\documentclass[fleqn,usenatbib]{mnras}
\usepackage{graphicx}
\usepackage{natbib}
\usepackage{ae,aecompl}
\usepackage{amsmath}
\usepackage{color}
\usepackage{comment}
\usepackage{rotating}
\usepackage{mathptmx}
\usepackage{txfonts}
\usepackage[T1]{fontenc}

\newcommand{\UFIG}{{\scshape UFig}}
\newcommand{\BALROG}{{\scshape Balrog}}
\newcommand{\altaffilmark}[1]{$^{#1}$}
\newcommand{\affil}[1]{{ #1}\\}

\usepackage{eso-pic}% http://ctan.org/pkg/eso-pic

\AddToShipoutPictureBG*{%
  \AtPageUpperLeft{%
    \hspace{0.75\paperwidth}%
    \raisebox{-3.5\baselineskip}{%
      \makebox[0pt][l]{\textnormal{DES 2015-0144}}
}}}%

\AddToShipoutPictureBG*{%
  \AtPageUpperLeft{%
    \hspace{0.75\paperwidth}%
    \raisebox{-4.5\baselineskip}{%
      \makebox[0pt][l]{\textnormal{FERMILAB-PUB-17-433-E}}
}}}%

\title[X-ray Cluster Red Sequence Luminosity Function]{Galaxies in X-ray Selected Clusters and Groups in Dark Energy Survey Data II: Hierarchical Bayesian Modeling of the Red-Sequence Galaxy Luminosity Function}
\author[DES Collaboration]{
\parbox{\textwidth}{
\Large
\mbox{Y.~Zhang$^{1}$ \thanks{Yuanyuan Zhang: \href{ynzhang@fna.gov}{ynzhang@fna.gov}}},
C.~J.~Miller\altaffilmark{2,3},
P.~Rooney\altaffilmark{4},
A.~Bermeo\altaffilmark{4},
A.~K.~Romer\altaffilmark{4},
C.~Vergara Cervantes\altaffilmark{4},
E.~S.~Rykoff\altaffilmark{5,6},
C.~Hennig\altaffilmark{7,8},
R.~Das\altaffilmark{3},
T.~McKay\altaffilmark{3},
J.~Song\altaffilmark{9},
H.~Wilcox\altaffilmark{10},
D.~Bacon\altaffilmark{10},
S.~L.~Bridle\altaffilmark{11},
C.~Collins\altaffilmark{12},
C.~Conselice\altaffilmark{13},
M.~Hilton\altaffilmark{14},
B.~Hoyle\altaffilmark{15},
S.~Kay\altaffilmark{16},
A.~R.~Liddle\altaffilmark{17},
R.~G.~Mann\altaffilmark{18},
N.~Mehrtens\altaffilmark{19},
J.~Mayers\altaffilmark{4},
R.~C.~Nichol\altaffilmark{10},
M.~Sahl\'en\altaffilmark{20},
J.~Stott\altaffilmark{21},
P.~T.~P.~Viana\altaffilmark{22,23},
R.~H.~Wechsler\altaffilmark{24,5,6},
T.~Abbott\altaffilmark{25},
F.~B.~Abdalla\altaffilmark{26,27},
S.~Allam\altaffilmark{1},
A.~Benoit-L{\'e}vy\altaffilmark{28,26,29},
D.~Brooks\altaffilmark{26},
E.~Buckley-Geer\altaffilmark{1},
D.~L.~Burke\altaffilmark{5,6},
A.~Carnero~Rosell\altaffilmark{30,31},
M.~Carrasco~Kind\altaffilmark{32,33},
J.~Carretero\altaffilmark{34},
F.~J.~Castander\altaffilmark{35},
M.~Crocce\altaffilmark{35},
C.~E.~Cunha\altaffilmark{5},
C.~B.~D'Andrea\altaffilmark{36},
L.~N.~da Costa\altaffilmark{30,31},
H.~T.~Diehl\altaffilmark{1},
J.~P.~Dietrich\altaffilmark{7,8},
T.~F.~Eifler\altaffilmark{37,38},
B.~Flaugher\altaffilmark{1},
P.~Fosalba\altaffilmark{35},
J.~Garc\'ia-Bellido\altaffilmark{39},
E.~Gaztanaga\altaffilmark{35},
D.~W.~Gerdes\altaffilmark{2,3},
D.~Gruen\altaffilmark{5,6},
R.~A.~Gruendl\altaffilmark{32,33},
J.~Gschwend\altaffilmark{30,31},
G.~Gutierrez\altaffilmark{1},
K.~Honscheid\altaffilmark{40,41},
D.~J.~James\altaffilmark{42},
T.~Jeltema\altaffilmark{43},
K.~Kuehn\altaffilmark{44},
N.~Kuropatkin\altaffilmark{1},
M.~Lima\altaffilmark{45,30},
H.~Lin\altaffilmark{1},
M.~A.~G.~Maia\altaffilmark{30,31},
M.~March\altaffilmark{36},
J.~L.~Marshall\altaffilmark{19},
P.~Melchior\altaffilmark{46},
F.~Menanteau\altaffilmark{32,33},
R.~Miquel\altaffilmark{47,34},
R.~L.~C.~Ogando\altaffilmark{30,31},
A.~A.~Plazas\altaffilmark{38},
E.~Sanchez\altaffilmark{48},
M.~Schubnell\altaffilmark{3},
I.~Sevilla-Noarbe\altaffilmark{48},
M.~Smith\altaffilmark{49},
M.~Soares-Santos\altaffilmark{1},
F.~Sobreira\altaffilmark{50,30},
E.~Suchyta\altaffilmark{51},
M.~E.~C.~Swanson\altaffilmark{33},
G.~Tarle\altaffilmark{3},
A.~R.~Walker\altaffilmark{25}
\begin{center} (DES Collaboration) \end{center}
\bigskip
\small{{\em Author affiliations are listed at the end of this paper.}}
}
}
\date{Accepted XXX. Received YYY; in original form ZZZ}

\pubyear{2019}

\begin{document}
\label{firstpage}
\pagerange{\pageref{firstpage}--\pageref{lastpage}}

\maketitle
\begin{abstract}
Using $\sim 100$ X-ray selected clusters in the Dark Energy Survey Science Verification data, we constrain the luminosity function (LF) of cluster red sequence galaxies as a function of redshift. This is the first homogeneous optical/X-ray sample large enough to constrain the evolution of the luminosity function simultaneously in redshift ($0.1<z<1.05$) and cluster mass ($13.5 \le \rm{log_{10}}(M_{200crit}) \sim< 15.0$). We pay particular attention to completeness issues and the detection limit of the galaxy sample. We then apply a hierarchical Bayesian model to fit the cluster galaxy LFs via a Schecter function, including its characteristic break ($m^*$) to a faint end power-law slope ($\alpha$). Our method enables us to avoid known issues in similar analyses based on stacking or binning the clusters.  We find weak and statistically insignificant ($\sim 1.9 \sigma$) evolution in the faint end slope $\alpha$ versus redshift. We also find no dependence in $\alpha$ or $m^*$ with the X-ray inferred cluster masses. However, the amplitude of the LF as a function of cluster mass is constrained to $\sim 20\%$ precision.  As a by-product of our algorithm, we utilize the correlation between the LF and cluster mass to provide an improved estimate of the individual cluster masses as well as the scatter in true mass given the X-ray inferred masses. This technique can be applied to a larger sample of X-ray or optically selected clusters from the Dark Energy Survey, significantly improving the sensitivity of the analysis. 
\end{abstract}
\begin{keywords}galaxies: evolution - galaxies: clusters: general 
\end{keywords}
%\email{\hspace{1em}$^\dagger$Yuanyuan Zhang}
%\maketitle

\section{Introduction}

 Galaxy clusters are special for both cosmology and astrophysics studies. As the structures that correspond to the massive end of halo mass function, they are sensitive probes of the $\Lambda$CDM cosmological model \citep[see reviews in][]{2011ARA&A..49..409A, 2013PhR...530...87W}. As the most massive virialized structures in the universe, they provide the sites for studying astrophysical processes in dense environments.

Galaxy clusters are known to harbor red sequence (RS) galaxies, so named because these galaxies rest on a tight relation in the color-magnitude space \citep{1992MNRAS.254..589B}. The phenomenon has been employed in finding clusters from optical data \citep[e.g.,][]{2000AJ....120.2148G, 2005AJ....130..968M, 2007ApJ...660..221K, 2016ApJS..224....1R, 2017arXiv170100818O} and developing cluster mass proxies \citep[e.g.,][]{2012ApJ...746..178R}. Red sequence galaxies also attract attention in astrophysics studies as they exhibit little star formation activity. Their formation and evolution provide clues to how quenching of galaxy star formation occurs in the cluster environment. 

It is well-established that the massive red sequence galaxies form at an early epoch \citep[e.g.,][]{2005ApJ...623L..85M, 2005ApJ...634L.129S, 2006ApJ...644..759M, 2008ApJ...684..905E, 2009A&A...504..331K, 2009ApJ...697..436H, 2010ApJ...716.1503P, 2011A&A...526A.133G, 2011MNRAS.410..280J, 2012MNRAS.423.3652G, 2013PASJ...65...17T}, but the formation of faint red sequence galaxies can be better characterized. The latter could be examined through inspecting the luminosity distribution of cluster galaxies, either with the dwarf-to-giant ratio approach \citep{2007MNRAS.374..809D}, or as adopted in this paper, with a luminosity function (LF) analysis. Results from these analyses are controversial to date, and have been extensively reviewed in literature \citep[e.g.,][]{2007ApJ...665..265F, 2009ApJ...690.1158C, 2014A&ARv..22...74B, 2015MNRAS.448....2W}. 

To summarize, a few studies have reported a deficit of faint red sequence galaxies with increasing redshift \citep{2007MNRAS.374..809D, 2007ApJ...661...95S, 2008ApJ...673..742G, 2009ApJ...700.1559R, 2010MNRAS.403.1274C, 2011MNRAS.414.2771D, 2015A&A...575A.116M, 2017arXiv170904484L}, indicating later formation of faint red sequence galaxies compared to the bright (and massive) ones. Yet, many other works observe little evolution in the red sequence luminosity distribution up to redshift $1.5$ \citep{2008MNRAS.386.1045A, 2009ApJ...690.1158C, 2013MNRAS.434.3469D, 2015MNRAS.450.1268D, 2016MNRAS.461.4517D, 2016MNRAS.457.2209C, 2017ApJ...848...37C, 2018A&A...613A..67S}, suggesting an early formation of both faint and bright red sequence galaxies. Differences in these results are hard to interpret given the different methods \citep[see the discussion in][]{2009ApJ...690.1158C}, sample selections and possible dependence on cluster mass \citep{2008ApJ...673..742G, 2009ApJ...699.1333H, 2015arXiv150907870L}, dynamical states \citep{2015MNRAS.448....2W, 2013MNRAS.434.3469D}, and whether or not the clusters are fossils \citep{2015A&A...581A..16Z}. Carrying out more detailed analyses, especially in the 0.5 to 1.0 redshift range, may help resolve the differences.

The luminosity distribution of cluster galaxies has also been modeled to connect galaxies with the underlying dark matter distribution. The luminosity function of galaxies in a halo/cluster of fixed mass, entitled the conditional luminosity function (CLF) in the literature \citep{2003MNRAS.339.1057Y}, statistically models how galaxies occupy dark matter halos. Modeling the Halo Occupation Distribution \citep[HOD,][]{2000MNRAS.318.1144P, 2002ApJ...575..587B, 2002MNRAS.329..246B} provides another popular yet closely-related approach.  Given a dark matter halo distribution, these models (HOD \& CLF) can be linked with several galaxy distribution and evolution properties \citep[e.g.,][]{2005A&A...433..415P, 2006MNRAS.365..842C, 2007A&A...464..451P, 2007ApJ...667..760Z, 2007MNRAS.376..841V, 2011ApJ...736...59Z, 2012ApJ...744..159L, 2013ApJ...771...30R}, including galaxy correlation functions \citep[e.g.,][]{1998ApJ...494....1J, 2000MNRAS.318.1144P,2000MNRAS.318..203S}, galaxy luminosity/stellar mass functions \citep[e.g.,][]{2009ApJ...695..900Y}, global star formation rate \citep[e.g.,][]{2013ApJ...770...57B} and galaxy-galaxy lensing signals \citep[e.g.,][]{2006MNRAS.368..715M}. 

Furthermore, LF \& HOD analyses improve our understanding of the cluster galaxy population. The number of cluster galaxies, especially the number of cluster red sequence galaxies, is a useful mass proxy for cluster abundance cosmology. Deep optical surveys like the Dark Energy Survey \citep[DES\footnote{https://www.darkenergysurvey.org},][]{DES2005} demand refined understanding of the evolution of cluster galaxies to $z=1.0$ \citep{2016arXiv161006890M} .

The Sloan Digital Sky Survey (SDSS\footnote{http://www.sdss.org}) has enabled detailed analysis of the cluster LFs (or CLFs) with the identification of tens of thousands of clusters to redshift 0.5 \citep{2008ApJ...676..248Y, 2009ApJ...699.1333H}. Above redshift 0.5, most studies have been performed with relatively small samples containing a handful of clusters or groups \citep{2008MNRAS.386.1045A, 2009ApJ...700.1559R, 2009ApJ...690.1158C, 2013MNRAS.434.3469D, 2015A&A...575A.116M, 2017MNRAS.465.4035D} and wide field surveys that are more sensitive than SDSS have just provided an opportunity to reinvigorate such analyses \citep{2018A&A...613A..67S}.

%@YyZfromKR: I don't think it is self explanatory that "SN data sets" were collected during SV phase, just not included in the SV-DR. Maybe it doesn't matter. And maybe you used post-SV data in the SN fields anyway in the analysis?

In this paper, we constrain the (conditional) red sequence luminosity function  (RSLF) with an X-ray selected cluster sample (details in Section~\ref{sec:cluster_sample}) detected in the DES Science Verification (DES-SV) data including the supernovae data sets collected during the same time. Clusters selected with the same approach are used in a cluster central galaxy study in \cite{2016ApJ...816...98Z}, but with an updated X-ray archival data set. The sample contains $\sim$ 100 clusters and groups in the mass range of $3\times 10^{13}$ $M_\odot$ to $2\times 10^{15}$ $M_\odot$, and the redshift range of 0.1 to 1.05. To date, it still represents a cluster sample that is complete to the highest redshift range discovered in DES, owing to the full depth data sets collected during DES-SV. As the clusters are not selected by their red sequence properties, studying RSLF with the sample is not subject to red sequence selection biases. Similar analyses can also be applied to SZ-selected clusters \citep[e.g., clusters discovered from the South Pole Telescope survey:][]{2015ApJS..216...27B, 2017MNRAS.467.4015H} and clusters selected from optical data.
Our paper focuses on cluster red members. The luminosity function of blue galaxies generally deviates from that of the red, but the red cluster members are easier to select photometrically due to the tightness of the color-magnitude relation.

%@YyZfromKR: when you say "acquire similar results with a stacking method" do you mean "acquire similar results to those obtainable using a stacking method" and maybe "but also a robust" could be "with the added advantage that robust"

The number of member galaxies in low mass clusters is often too low to study LFs for individual systems. 
It is a common approach to stack the member galaxy luminosity distributions for an ensemble of clusters \citep[e.g., ][]{2009ApJ...695..900Y, 2009ApJ...699.1333H}.
In this paper, we develop a hierarchical Bayesian modeling technique.  The method allows us to acquire similar results to a stacking method, with the added benefits of robust uncertainty estimation and simultaneous quantification of the possible mass dependence and redshift evolution effects. In the rest of the paper, we first introduce our data sets in Section~\ref{sec:data} and then describe the methods in Section~\ref{sec:methods}. The results are presented in Section~\ref{sec:results}. Discussions of the methods and results as well as a summary of the paper are presented in Section~\ref{sec:discussion}.  

\section{Data}
\label{sec:data}

\subsection{Dark Energy Survey Science Verification Data}

We use the DES Science Verification (DES-SV) data taken in late 2012 and early 2013.  
The DES collaboration collected this data set with the newly mounted Dark Energy Camera \citep[DECam, ][]{2015arXiv150402900F} for science verification purposes before the main survey began \citep[for details on DES Year 1 operations, see][]{2014SPIE.9149E..0VD}. In total, the data set covers $\sim$ 400 deg$^2$ of the sky. For about 200 deg$^2$, data are available\footnote{http://des.ncsa.illinois.edu/releases/sva1} in all of the $g$, $r$, $i$, $z$ and $Y$ bands, and the total exposure time in each band fulfills DES full depth requirement \citep[23 to 24 mag in $i$ and 22 to 23 mag in $z$, see more details in][]{2014MNRAS.445.1482S}. A pilot supernovae survey \citep[see][for an overview]{2015MNRAS.449.1215P} of 30 deg$^2$ sky in $g$, $r$, $i$, $z$ was conducted at the same time, reaching deeper depth after image coaddition ($\sim$ 25 mag in $i$ and $\sim$24 mag in $z$). 

The DES-SV data are processed with the official DES data reduction pipeline \citep[][]{2011arXiv1109.6741S, 2012SPIE.8451E..0DM}. In this pipeline, single exposure images are assessed, detrended, calibrated and coadded.  The coadded images are then fed to the SExtractor software \citep{1996A&AS..117..393B, 2011ASPC..442..435B}  for object detection and photometry measurement. 

\subsection{The DES Photometric Data}
\label{sec:sva1_gold}

We use a DES value-added catalog, the ``gold'' data set  \citep[see the review in][]{2016ApJS..224....1R, 2017arXiv170801531D}\footnote{https://des.ncsa.illinois.edu/releases/sva1}, based on catalogs produced from the SExtractor software. The detection threshold is set at 1.5$\sigma$ ( {\tt DETECT\_THRESH = 1.5}) with the default SExtractor convolution filter. The minimum detection area is set at 6 pixels\footnote{DECam pixel scale 0.263''} ({\tt DETECT\_MINAREA =6}). The SExtractor runs were performed in dual mode, using the linear addition of $r$, $i$ and $z$ band images as the detection image.

The ``gold'' data set is subsequently derived with the initial detections, keeping only regions that are available in all of the $g$, $r$, $i$, $z$ bands. Regions with a high density of outlier colors due to the impact of scattered light, satellite or  airplane trails,  and regions with low density of galaxies near the edge of the survey are removed. Objects near bright stars selected from the Two Micron All Sky Survey \citep[2MASS][]{2006AJ....131.1163S} are masked. The masking radius scales with stellar brightness in $J$ as   $R_\mathrm{mask} =150-10 J$  (arcseconds) with a maximum of 120 arcseconds \citep{2016MNRAS.460.2245J, 2016ApJS..224....1R}. Stars of nominal masking radii less than 30 arcseconds are not masked to avoid excessive masking. 
Coverage of the sample is recorded with the HEALPix\footnote{http://healpix.sourceforge.net} software \citep{2005ApJ...622..759G} gridded by $N=4096$.
Photometry are re-calibrated and extinction-corrected using the Stellar Locus Regression technique \citep[SLR:][]{2014MNRAS.439...28K}. 

We make use of the SExtractor Kron magnitudes \citep[mag\_auto, ][]{1980ApJS...43..305K} for all detected objects. Since the SExtractor run was performed in dual mode, the Kron aperture and the centroid for different filters are the same, which are determined from the detection images. The luminosity functions are derived with DES $z-$band photometry, based on objects > 5$\sigma$ (which corresponds to magerr\_auto\_z$<$ $2.5/ \mathrm{ln} 10/5$=0.218mag).

We derive completeness limits for the selected > 5$\sigma$ objects. Details of the completeness analyses are provided in Appendix~\ref{sec:completeness}. In general, the completeness limits are $\sim$0.5mag brighter than the sample's 10$\sigma$ depth magnitudes. The  selected > 5$\sigma$  objects are $>$99.8\% complete above the limits. Because of this high completeness level, we do not correct for incompleteness in this paper. 
 
\subsection{The XCS-SV cluster sample}
\label{sec:cluster_sample}

\begin{figure}
\includegraphics[width=0.5\textwidth]{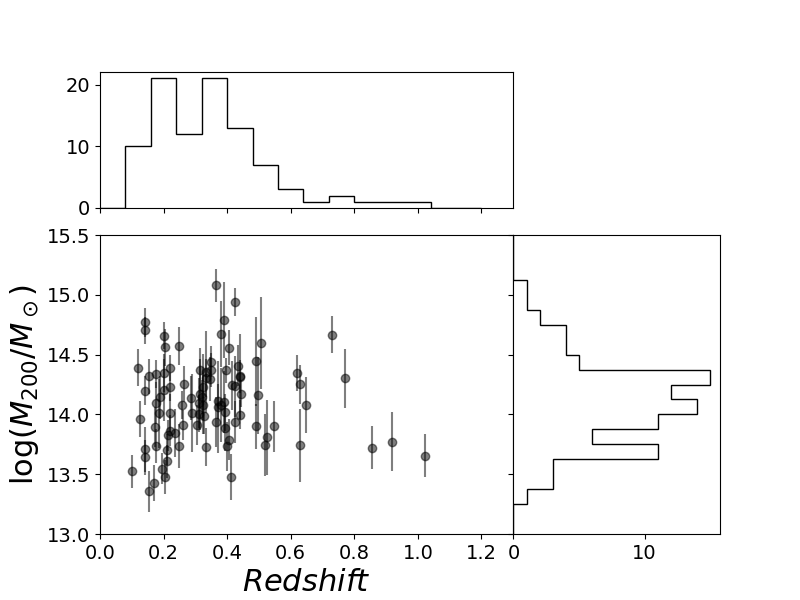}
\caption{The XCS-SV clusters: redshifts, masses, and mass uncertainties. The upper and right histograms respectively show the cluster redshift and mass distribution.}
\label{fig:redshift_mass}
\end{figure}

The XCS-SV cluster sample is a product from the {\it XMM} Cluster Survey \citep{2011MNRAS.418...14L, 2012MNRAS.423.1024M, 2013AN....334..462V}, which searches for galaxy cluster candidates (extended X-ray emissions) in the {\it XMM-Newton} archival data. The X-ray selected cluster candidates (about 300 in number) are later confirmed with the DES-SV optical images, and have their photometric redshifts estimated using the DES-SV photometric data set. The XCS-SV sample contains galaxy groups, low mass clusters and clusters as massive as $10^{15}\mathrm{M_\odot}$ to beyond redshift 1. Selection and confirmation methods of the sample, as well as the cluster photometric redshift measurements are reviewed in \citet[][henceforth referenced as Z16]{2016ApJ...816...98Z}. The sample used in this paper are expanded from that in Z16 after finalizing the input X-ray data. We make use of only the clusters of which the mass uncertainties, derived from the X-ray temperature measurements, are less than 0.4 dex.

Since this paper evaluates luminosity function with the $z$-band photometry, we eliminate clusters above redshift 1.05 for which the rest-frame 4000\AA~break of RS galaxies have shifted out of DES $z-$band coverage (sensitive to $\sim$8500~\AA ). We only use clusters located in DES-SV regions with the analysis magnitude ranges (above characteristic magnitude $+ $ 2 mag) above the completeness limits (Section~\ref{sec:sva1_gold}). The paper works with 93 clusters in total, which are listed in Appendex~\ref{sec:sample_table}, Table~\ref{tbl:clusters}. In Figure~\ref{fig:redshift_mass}, we show the redshifts, masses, and mass uncertainties of the analyzed clusters.

The cluster masses and uncertainties are derived from X-ray temperature based on a literature $T_X - M$ relation \citep{2013ApJ...778...74K} (see details also in Z16). $R_{200}$ is derived from $M_{200}$.

\subsection{Red Sequence Galaxy Selection}
\label{sec:select_red}

\begin{figure}
\includegraphics[width=0.5\textwidth]{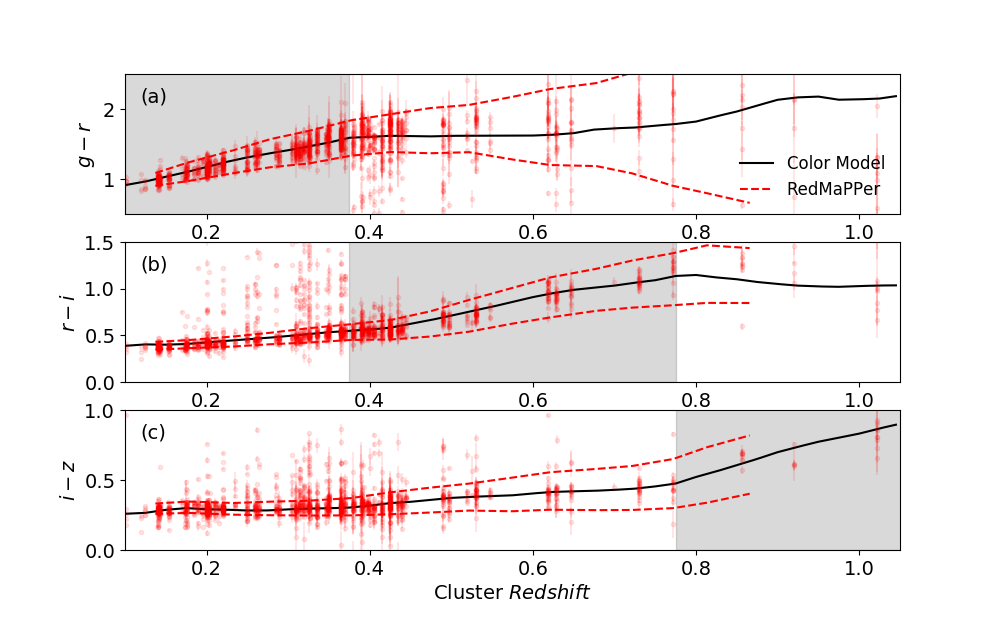}
\caption{Observer-frame  $g-r$ (panel a), $r-i$ (panel b) and $i-z$ (panel c) colors of the cluster red sequence candidates (red data points) and the red sequence model (black solid lines). Note that the color distributions of cluster foreground/background objects are not subtracted. We also show the 2$\sigma$ color ranges of red sequence cluster members (member probability $>$ 50\% ) from the redMaPPer DES-SV cluster sample  \citep{2016ApJS..224....1R} for comparison, which appear to agree with our color models.}
\label{fig:colormodel}
\end{figure}

The definition of cluster member galaxies in projected datasets is a difficult challenge. Our method is based on simple color cuts around the cluster red sequence \citep{2007MNRAS.374..809D, 2007ApJ...661...95S, 2008ApJ...673..742G, 2009ApJ...690.1158C, 2015A&A...575A.116M}. To account for the shifting of the 4000~\AA~break, we select red sequence galaxies according to $g-r$ color at $z< 0.375$, $r-i$ color at $0.375<=z<0.775$ and $i-z$ color at $0.775<=z<1.1$

For a cluster at redshift $z$, we first apply K-corrections \citep{2007AJ....133..734B} to all the objects in the cluster field. These objects are  band-shifted to a reference redshift (depending on the color choice), assuming the cluster redshift to be their original redshifts. We compare the corrected colors to a model color  with the following standard:

\begin{equation}
\begin{split}
| (g-r)_{z=0.25}  - (g-r)_\text{model at $z$=0.25} | &< \sqrt{ \delta^2_{g-r} +\Delta_{g-r}^2}, ~\text{or} \\
| (r-i)_{z=0.55} - (r-i)_\text{model at $z$=0.55} | &< \sqrt{ \delta^2_{r-i} +\Delta_{r-i}^2} , ~\text{or} \\
| (i-z)_{z=0.9} - (r-i)_\text{model at $z$=0.9} | &< \sqrt{ \delta^2_{i-z} +\Delta_{i-z}^2}.
\end{split}
\end{equation}
In these equations, the model colors ($g-r$, $r-i$, or $i-z$, details explained below) are the mid-points of a selection window at a reference redshift. $\delta_{g-r}$, $\delta_{r-i}$ and $\delta_{i-z}$are the photometry uncertainties. $\Delta_{g-r}$, $\Delta_{r-i}$ and $\Delta_{i-z}$ are the widths of the selection windows. 

We set $\Delta_{g-r}$ to be 0.2 $\mathrm{mag}$. The clipping width is chosen to be larger than the combination of the intrinsic scatter and the slope of red sequence color-magnitude relations, while avoiding a significant amount of blue galaxies. $\Delta_{r-i}$ is adjusted to be 0.15 through matching the number of selected cluster galaxies (after background subtraction, see Section~\ref{sec:histogram} for details) to fiducial $g-r$ selections at $0.3\le z<0.5$. $\Delta_{i-z}$ is adjusted to be 0.12 through matching the number of selected cluster galaxies (after background subtraction, see Section~\ref{sec:histogram} for details) to fiducial $r-i$ selections at $z\ge 0.7$.

The model colors  of $g-r$ at $z=0.25$, $r-i$ at $z=0.55$ and $i-z$ at $z=0.9$ are based on a simple stellar population template from \cite{2003MNRAS.344.1000B}, assuming a single star burst of metallicity $Z=0.008$ at $z=3.0$, computed with the EZGal package \footnote{http://www.baryons.org/ezgal/} \citep{2012PASP..124..606M}.  In Figure~\ref{fig:colormodel}, we show the red sequence model, over-plotting the observer frame colors of the selected objects. Overall, the colors of the selected RS candidates match template well. The template also matches the colors of cluster red sequence defined by the RedMaPPer method \citep{2016ApJS..224....1R}. 

For RS candidates selected with the above criteria, we employ a statistical background subtraction approach (see details in Section~\ref{sec:methods}) to eliminate background objects, which on average constitute 50\% of the cluster region galaxies brighter than $m^*$+2 mag. 

The performance of star-galaxy classifiers applied to the DES SVA1 ``gold" sample (Section~\ref{sec:sva1_gold}) depends on the object's apparent magnitude. The classifiers become unstable for objects fainter than $\sim$22 mag in the $z-$band. Since it is possible to eliminate the stellar contamination with the background subtraction procedure (we estimate the background object --stellar and galactic-- densities locally for each cluster), we do not attempt to separate stars and galaxies among the RS candidates (above 21 mag in $z$, stars make up $\sim$10\% of the sample). We nevertheless refer to all objects as ``galaxies''.

\section{Methods}
\label{sec:methods}

The main results in this paper are derived with a hierarchical Bayesian method \citep[application examples to cosmology can be found in ][]{2010bmic.book..245L}. We constrain the RSLF with a single Schechter function \citep{1976ApJ...203..297S}  to the magnitude limit of $m_* +2$ mag, and simultaneously model the mass and redshift dependence of the parameters (Section~\ref{sec: bayes}: a hierarchical Bayesian method). To test the method, we compare the constraints to results derived  from stacking cluster galaxy number counts in luminosity bins (Section~\ref{sec:histogram}: alternative histogram method). 

Generally, the input to both methods includes the observed magnitudes, $\{m_i\}$,  of objects inside clusters or in a "field" region ($m_i$ is the apparent magnitude of the $i$th object). We define the cluster region as enclosed within 0.6 $R_{200}$ of the cluster centers (X-ray centers). The contrast between cluster and background object densities is large with this  choice (excess cluster object density to background object density about 1:1 for most of the clusters throughout the DES-SV depth), and the amount of retained cluster galaxies is reasonable. We choose the field region to be annular, centered on the cluster, with the inner and outer radii being 3 $R_{200}$ and 8 $R_{200}$ respectively. The choice helps eliminating RSLF contributions from cluster-correlated large scale structures along the line-of-sight. The cluster central galaxies selected according to the criteria in Z16 are eliminated from the analysis. Central galaxies are known to be outliers to a Schechter function distribution. Their properties and halo occupation statistics are investigated in Z16.

The area of these regions are traced with randomly generated locations that have uniform surface density across the ``gold" sample footprint, i.e., a sample of ``random points''.  For each cluster, we generate $\sim 1.5$ million random points within 10 $R_{200}$. The number density is high enough that the resulting uncertainty is negligible ($\sim 1\%$ in the luminosity distribution measurements). We ignore the uncertainties from using random points.

\subsection{A Hierarchical Bayesian Method}
\label{sec: bayes}

Given a model with a set of parameters $\Omega$ that describes the distribution of observables, Bayesian theory provides a framework for inferring $\Omega$ with a set of observed quantities $\{x\}$. In this sub-section, we describe the methods developed in this framework.  

Denoting the probability of observing  $\{x\}$ in model $\Omega$ to be $P(\{x\} | \Omega) $, and the prior knowledge about  the model parameters to be $P(\Omega)$, after observations of $\{x\}$, the Bayes' theorem updates the  knowledge about model parameters, namely the posterior distribution, to be:
\begin{equation}
P(\Omega | \{x\} )  \propto P(\{x\} | \Omega) P(\Omega).
\end{equation}
The above equation uses a proportional sign instead of an equal sign as a probability function needs to be normalized to 1. The normalization factor is un-interesting when the posterior probability is sampled with Markov Chain Monte Carlo.

In our application, the observables include the observed magnitudes of objects in the cluster or field region. A major component of our model is the Schechter function. The parameters of the Schechter function vary for clusters of different masses and redshifts. Our model, called the {\it hierarchical} model, assumes redshift and mass dependences for the faint end slope and the characteristic magnitude.  For the parameter priors $P(\Omega)$, we assume them to be flat for most of the parameters excluding a couple. The prior distributions are noted later when we introduce the parameters.

\subsubsection{Basic Components of the Model}
\label{sec:models}

For {\bf one cluster galaxy}, we assume that the probability of observing it with magnitude $m$ follows a Schechter function: 
\begin{equation}
f(x) = \psi_f  (\mathrm{0.4 ln 10} )10^{0.4(m^*-x)(\alpha+1)} \mathrm{exp}(- 10^{0.4(m^*-x)})\label{eq:fx}
\end{equation}
In this equation, $\psi_f$ is the normalization parameter that normalizes $f(x)$ to 1. $\alpha$ and $m^*$ are the faint end slope and the characteristic magnitude,  treated as free parameters of the model.

For one object in the {\bf cluster region}, it can be either a {\it cluster galaxy} or a {\it field object}. For a {\bf field object}, we denote the probability of observing it with magnitude $m$ to be $g(m)$. $g(m)$ is approximated with a normalized histogram of the object magnitude distribution in the field region.

The probability of observing one object in the cluster region is the combination of observing it as a {\it field} object {\bf and} observing it as a {\it cluster} galaxy. The probability writes
\begin{equation}
h(m) = \psi_h [N_\mathrm{cl} f(m) + N_\mathrm{bg} g(m)]
\label{eq:hx1}
\end{equation}
In this equation, $N_\mathrm{cl}$ is the number of cluster galaxies in the cluster region, and $N_\mathrm{bg} $ is the number of field galaxies in the cluster region. Again, there exists a normalization factor  $\psi_h$ that normalizes the probability function to 1.  

We treat the sum of $N_\mathrm{bg} $ and $N_\mathrm{cl}$ as a Poisson distribution. The expected value of $N_\mathrm{bg} $ can be extrapolated from the field region and the area ratio between the cluster and the field regions. Equation~\ref{eq:hx1} introduces {\it one} free parameter, $N_\mathrm{cl}$, which controls the relative density between cluster and field galaxies in the cluster region. $N_\mathrm{cl}$ can be further related to the amplitude of the Schechter function, $\phi^*$ (in unit of total galaxy count), as  the integration of the Schechter function over the interested magnitude range, written as
\begin{equation}
\begin{split}
N_\mathrm{cl} &= \int  \frac{\phi^* f(m)} {\psi_f} \mathrm{d} m \\
& = \frac{\phi^* } {\psi_f}  \int  \ f(m) \mathrm{d} m .\\
\end{split}
\end{equation}

Thus far, the free parameters in our models are $\alpha$, $m^*$ from Equation~\ref{eq:fx} and $\phi^*$. Note that, in this section, we only perform analyses with galaxies brighter than the completeness magnitude limit (galaxies are considered to be more than 99.8\% complete throughout the analyzed magnitude range, according to Section~\ref{sec:completeness}).

We constrain $\phi^*$ with the number count of observed objects in the cluster region ($N$), assuming a Poisson distribution:

\begin{equation}
N \sim \mathcal{P}oisson(N_\mathrm{cl}+N_\mathrm{bg}).
\end{equation}
The log-likelihood is explicitly written as:
\begin{equation}
\mathrm{log} \mathcal{P}(N)  \propto N\mathrm{log}(N_\mathrm{cl}+N_\mathrm{bg}) - (N_\mathrm{cl}+N_\mathrm{bg}).
\end{equation}

For one cluster, we take the observables to be the observed magnitudes of cluster region objects, $\{m_i\}$, the object number count and $N$ and the background object number count. $N_\mathrm{bg}$ is treated as a known quantity. The log-likelihood of observing these quantities is:

\begin{equation}
\begin{split}
\mathrm{log} \mathcal{P} & (\{m_i\},  N | \alpha, m^*, \phi^*) \\
 &\propto \mathrm{log}\mathcal{P}(N| \phi^*, \alpha, m^*) + \sum_{i} \mathrm{log}\mathcal{P} (\{m_i\}| \alpha, m^*, \phi^*)  \\
&\propto \mathrm{log} \mathcal{P}(N) + \sum_{i} \mathrm{log} h(m_i) .
\end{split}
\end{equation}

\subsubsection{Hierarchical Model}
\label{sec:dependence}

The Bayesian approach makes it possible to add dependences to $\alpha$ and $m^*$ . We rewrite $\alpha$ and $m^*$ with redshift or mass dependences:
\begin{equation}
\begin{split}
\alpha_j & = A_\alpha \mathrm{log}(1+z_j) +B_\alpha (\mathrm{log} M_{\mathrm{model}, j}-14) +C_\alpha \\
m^*_{z=0.4, j} & = B_m  (\mathrm{log} M_{\mathrm{model}, j}-14) +C_m.
\label{eq:z_mass_dependence}
\end{split}
\end{equation}
Here, we distinguish between true and observed $M_{200}$ of clusters. $\mathrm{log} M_{\mathrm{model}, j}$ represents the true $M_{200}$ mass of the $j$th cluster, while we use $\mathrm{log} M_{\mathrm{obs}, j}$ to represent the $M_{200}$ mass derived from X-ray temperature for the $j$th cluster. $\mathrm{log}  M_{\mathrm{model}, j}$ for different clusters are treated as free parameters in the analysis, but we use observational constraints on $\mathrm{log}  M_{200}$ from X-ray data  as priors (Gaussian distributions): $\mathrm{log}  M_{\mathrm{model}, j}$ $\sim$ $\mathcal{N}(\mathrm{log}  M_{\mathrm{obs}, j}, \sigma_M^2)$. $\sigma_M$ is the measurement uncertainty (including the intrinsic scatter and statistical uncertainties) of $\mathrm{log}  M_{\mathrm{obs}, j}$ from X-ray data. The assumption about  $\mathrm{log} M_{\mathrm{model}, j}$ allows us to incorporate mass uncertainties into the analysis. Furthermore, we constrain $m^*$ at $z=0.4$ (the mean and median redshifts of the sample are 0.33 and 0.35 respectively) to be consistent with the redshift cut in the alternative method in Section~\ref{sec:histogram}. For each cluster, we extrapolate the $m^*$ at its observed $z$ from $z=0.4$ assuming a simple stellar population from \cite{2003MNRAS.344.1000B} with a single star burst of metallicity $Z=0.008$ at $z=3.0$ (the red sequence galaxy template used in Section~\ref{sec:select_red}).

$\phi^*$ for each cluster is constrained separately. 
We assume a Gaussian prior distribution of $\{ \mathrm{log} \phi^*_j\}$ given the values predicted by the relation: $\phi^*_j \sim$  $\mathcal{N}(\mathrm{log} \phi^*_{\mathrm{mean}, j}, \sigma_{ \mathrm{log}\phi }^2)$. 
$\sigma_{ \mathrm{log} \phi }$ is the intrinsic scatter of the relation, fixed at 0.5 \footnote{Allowing the parameter to vary gives a scatter of $\sim$0.2 to 0.3, and therefore we decided to set a value conservatively larger to avoid over-constraining the $\sigma_{ \mathrm{log} \phi }$ parameters} to reduce the number of free parameters. 
We further assume a power law relation between $ M_{\mathrm{model}, j}$ and $\phi^*_{\mathrm{mean}, j}$:
\begin{equation}
\mathrm{log}\phi^*_{\mathrm{mean}, j}= B_\phi \times \mathrm{log} M_{\mathrm{model}, j}+C_\phi.
\label{eq:phi_mass_dependence}
\end{equation}
The log likelihood of having $ \phi^*_j$ given $M_{\mathrm{model}, j} $ writes:
\begin{equation}
\begin{split}
g_j (\phi^*_j) \propto -\frac{ \big(\phi^*_j-( B_\phi \times \mathrm{log} M_{\mathrm{model}, j}+C_\phi)\big)^2}{2\sigma_{\mathrm{log}\phi}^2}
\end{split}
\end{equation}

\begin{figure}
\includegraphics[width=0.5\textwidth]{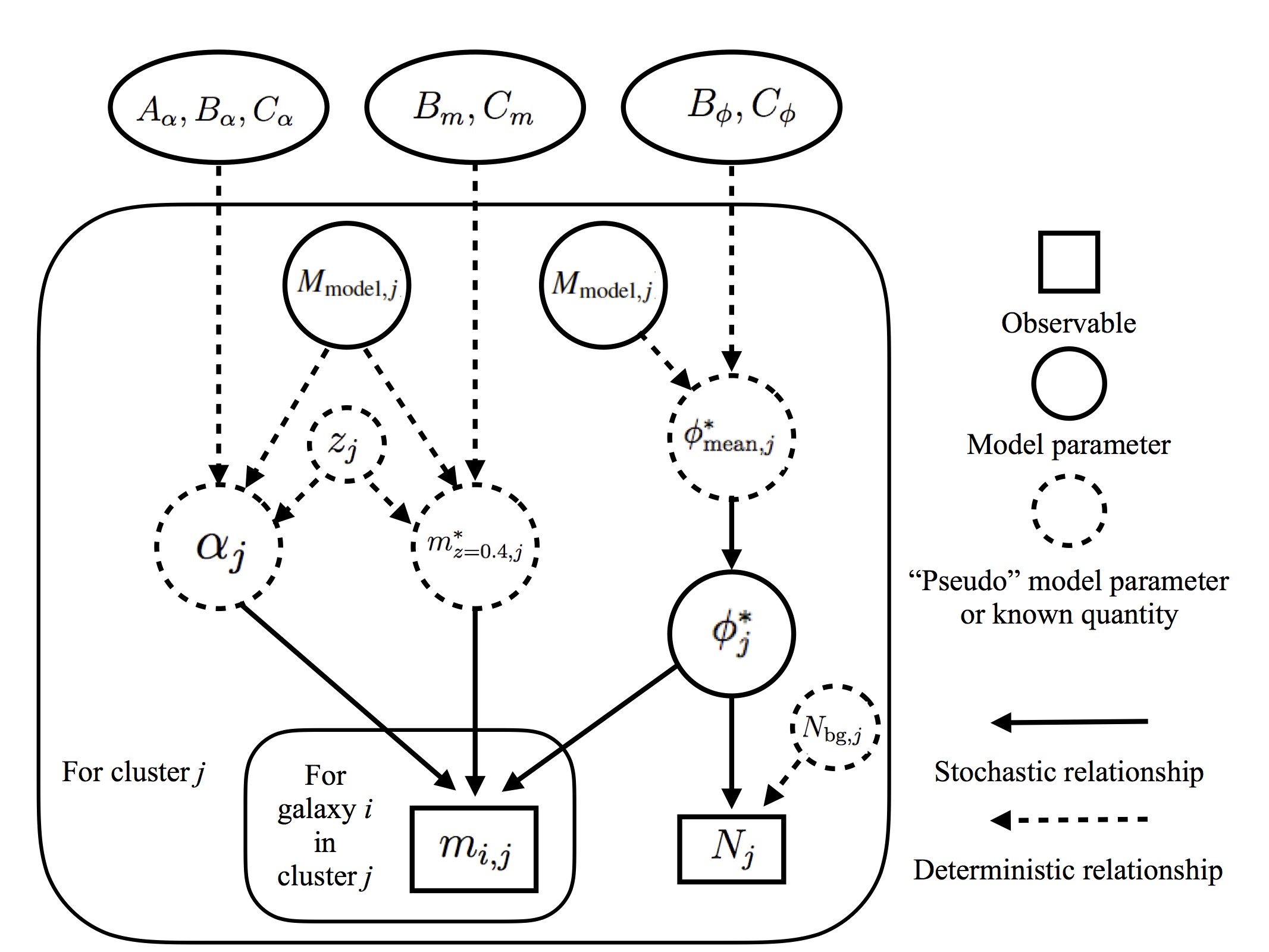}
\caption{Schematic diagram of the hierarchical Bayesian method, as described in Section~\ref{sec: bayes}. Note that Schechter function parameters like $\alpha_j$, $m*_{z=0.4, j}$ and $\phi^*_j$ are not directly constrained in the model. Such ``parameters'' (called pseudo parameters in the diagram),  as well as known quantities are indicated by dashed line circles.}
\label{fig:flowchart}
\end{figure}

The free parameters of this model are $A_\alpha$, $B_\alpha$, $C_\alpha$, $B_m$, $C_m$, $B_\phi$, $C_\phi$, $\{ \phi^*_j\}$ and $\{ M_{\mathrm{model}, j} \}$. The observed quantities are $\{m_{i, j}\}$ and $\{N_j\}$ of all clusters. $\{\mathrm{log} M_{\mathrm{obs}, j}\}$ are treated as priors for $\{\mathrm{log} M_{\mathrm{model}, j}\}$. $\{z_j\}$ as well as $N_{\mathrm{bg}, j} $ are treated as known quantities for each of the clusters. We summarize the model dependences with a schematic diagram in Figure~\ref{fig:flowchart}. The log-likelihood of observing these quantities is:

\begin{equation}
\begin{split}
\mathrm{log} \mathcal{L} & (\{m_{i, j}\},  \{N_j\} |A_\alpha, B_\alpha, C_\alpha, B_m, C_m, B_\phi, C_\phi, \{ \phi^*_j\}, \{ M_{\mathrm{model}, j} \} ) \\
&=\mathrm{log} \mathcal{L}  (\{m_{i, j}\},  \{N_j\} | \alpha_j, m^*_j, \{ \phi^*_j\}) + \mathrm{log} \mathcal{L} ( \{ \phi^*_j\} | \{ M_{\mathrm{model}, j} \})\\
 &\propto \sum_{j} [\mathrm{log}\mathcal{P}(N_j| \phi^*_j, \alpha_j, m^*_j) + \sum_{i} \mathrm{log}\mathcal{P} (\{m_{i, j}\}| \alpha_j, m^*_j, \phi^*_j) ]\\
 &+ \sum_{j} \mathrm{log} \mathcal{L} ( \phi^*_j |  M_{\mathrm{model}, j} ) \\
&\propto \sum_{j} [ \mathrm{log} \mathcal{P}_j(N_j) + \sum_{i}  \mathrm{log} h_j(m_i, j) + g_j (\phi^*_j)] .
\end{split}
\end{equation}

Finally, the parameter posterior likelihood is
\begin{equation}
\begin{split}
\mathrm{log} \mathcal{L} & (A_\alpha, B_\alpha, C_\alpha, B_m, C_m, B_\phi, C_\phi, \{ \phi^*_j\}, \{ M_{\mathrm{model}, j} \} | \{m_{i, j}\},  \{N_j\} ) \\
&=\mathrm{log} \mathcal{L}  (\{m_{i, j}\},  \{N_j\} | \alpha_j, m^*_j, \{ \phi^*_j\}) + \mathrm{log} \mathcal{L} ( \{ \phi^*_j\} | \{ M_{\mathrm{model}, j} \})\\
 &\propto \sum_{j} [\mathrm{log}\mathcal{P}(N_j| \phi^*_j, \alpha_j, m^*_j) + \sum_{i} \mathrm{log}\mathcal{P} (\{m_{i, j}\}| \alpha_j, m^*_j, \phi^*_j) ]\\
 &+ \sum_{j} \mathrm{log} \mathcal{L} ( \phi^*_j |  M_{\mathrm{model}, j} )\\& + \mathrm{log} \mathcal{L} _\mathrm{prior} (A_\alpha, B_\alpha, C_\alpha, B_m, C_m, B_\phi, C_\phi, \{ \phi^*_j\}, \{ M_{\mathrm{model}, j} \}) \\
&\propto \sum_{j} [ \mathrm{log} \mathcal{P}_j(N_j) + \sum_{i}  \mathrm{log} h_j(m_i, j) + g_j (\phi^*_j)] \\
& + \mathrm{log} \mathcal{L} _\mathrm{prior} (A_\alpha, B_\alpha, C_\alpha, B_m, C_m, B_\phi, C_\phi, \{ \phi^*_j\}, \{ M_{\mathrm{model}, j} \}).
\end{split}
\end{equation}

We assume flat priors for most of the model parameters except $C_m$ and $\phi_j$.  For $C_m$, we assume a Gaussian distribution as the prior, with  the measurement from  \cite{2009ApJ...699.1333H} as the mean, and 1 mag$^2$ as the variance. These priors are listed in Table~\ref{tbl:alpha_mstar}. Sampling from the parameter posterior likelihood is performed with the {\sc PyMC} package \citep[][]{2015ascl.soft06005F}.

\subsection{Alternative Histogram Method}
\label{sec:histogram}

\begin{figure}
\includegraphics[width=0.5\textwidth]{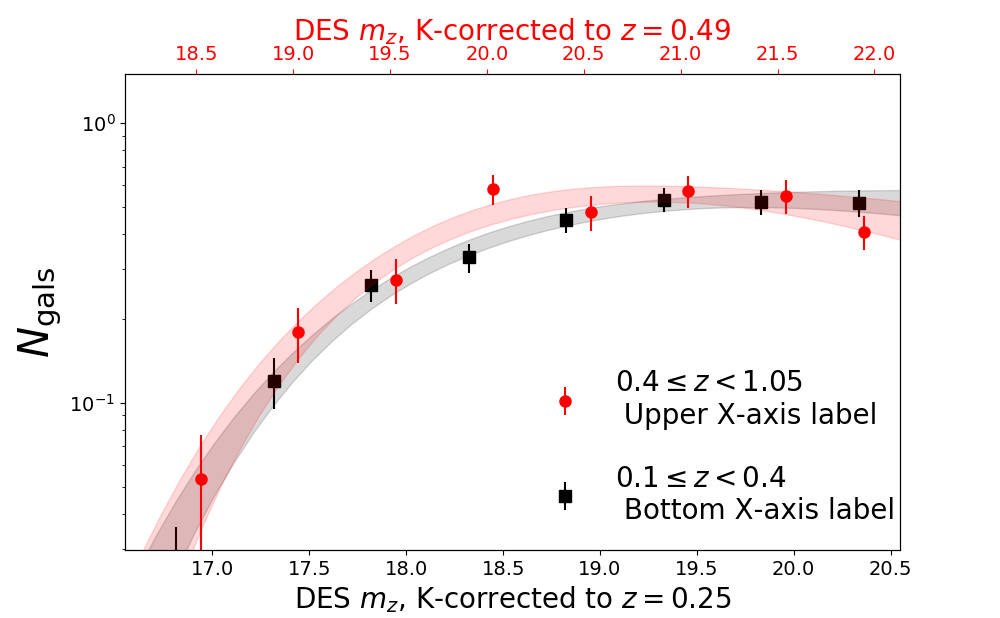}
\caption{RSLFs derived in two redshift bins display a possible redshift evolution effect. Uncertainties with the data points are estimated through assuming Poisson distributions. The shaded bands show the fitted Schechter functions including $1\sigma$ fitting uncertainties (with the method from Section~\ref{sec:histogram}). Note that the data points have been rebinned from the input to the fitting method.}
\label{fig:hist}
\end{figure}

\begin{figure}
\includegraphics[width=0.48\textwidth]{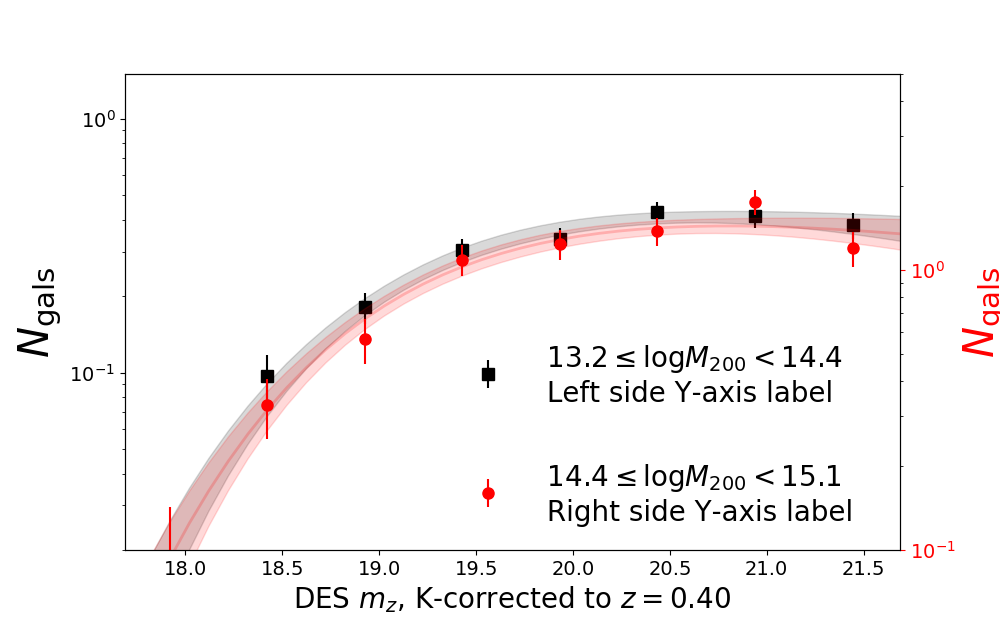}
\caption{RSLFs derived in two cluster mass bins appear to be consistent. Uncertainties with the data points are estimated through assuming Poisson distributions. The shaded bands show the fitted Schechter functions including the 1 $\sigma$ fitting uncertainties (with the method from Section~\ref{sec:histogram}). Note that the data points have been rebinned from the input to the fitting method.}
\label{fig:hist_mass}
\end{figure}

We develop a separate method to test the fore-mentioned technique. This method starts with counting galaxies in magnitude bins. We use 150 bins from 15 mag to 30 mag spaced by 0.1mag. We do not see change of the results when adjusting the bin size from 0.2 mag to 0.05 mag. 

The histogram counting is performed for the cluster region, $N(m)$, and the field region, $N(m)_\mathrm{background}$. To estimate the contribution of field galaxies to the cluster histogram, we weight the number count of objects in the field region, with the random number ratio:
\begin{equation}
N_\mathrm{bg}(m)=N(m)_\mathrm{background} \times  \frac{N_\mathrm{random, ~cluster}}{N_\mathrm{random, ~background}}. 
\end{equation}

We add up the histograms of clusters binned by redshift or cluster mass\footnote{ Two clusters are further eliminated from the 93 cluster sample because they are severly masked and therefore do not reliably contribute to the stacked histograms. }, and also record the number count of clusters in each magnitude bin, $C(m)$.  During the summing process, we shift $m$ by the apparent magnitude difference beween the cluster redshift and a reference redshift (depending on the cluster redshift and mass binning) of a simple passively-evolving stellar population from \cite{2003MNRAS.344.1000B} with a single star burst of metallicity $Z=0.008$ at $z=3.0$ (the same red sequence galaxy template used in Section~\ref{sec:select_red} and ~\ref{sec:dependence}). The histograms are then averaged for both the cluster region and the field region to obtain $\bar{N}(m)$  and $\bar{N}_\mathrm{bg}(m)$. Subtracting $\bar{N}_\mathrm{bg}(m)$ from $\bar{N}(m)$ yields the luminosity distribution of cluster galaxies (Figure~\ref{fig:hist} in redshift bins and Figure~\ref{fig:hist_mass} in mass bins).

We assume a Schechter function distribution for cluster galaxies:
\begin{equation}
S( m)= \phi (\mathrm{0.4 ln 10} )10^{0.4(m^*-m)(\alpha+1)} \mathrm{exp}(- 10^{0.4(m^*-m)} ), \label{eq:sm}
\end{equation}
therefore the expected number of galaxies in each magntiude bin in the cluster region is
\begin{equation}
E( m)=S( m)+N_\mathrm{bg}(m).
\end{equation}

Assuming Poisson distributions for the number of galaxies in each bin, we sample from the following likelihood:
\begin{equation}
\mathrm{log} \mathcal{L}  \propto \sum_{m} \bar{N}(m)C(m) \mathrm{log} [E( m)C(m) ] - E( m)C(m)  .
\end{equation}

Sampling from the likelihood is performed with the {\sc emcee} package \citep[][]{2013PASP..125..306F}.
\section{Results}
\label{sec:results}

\begin{figure*}
\includegraphics[width=1.0\textwidth]{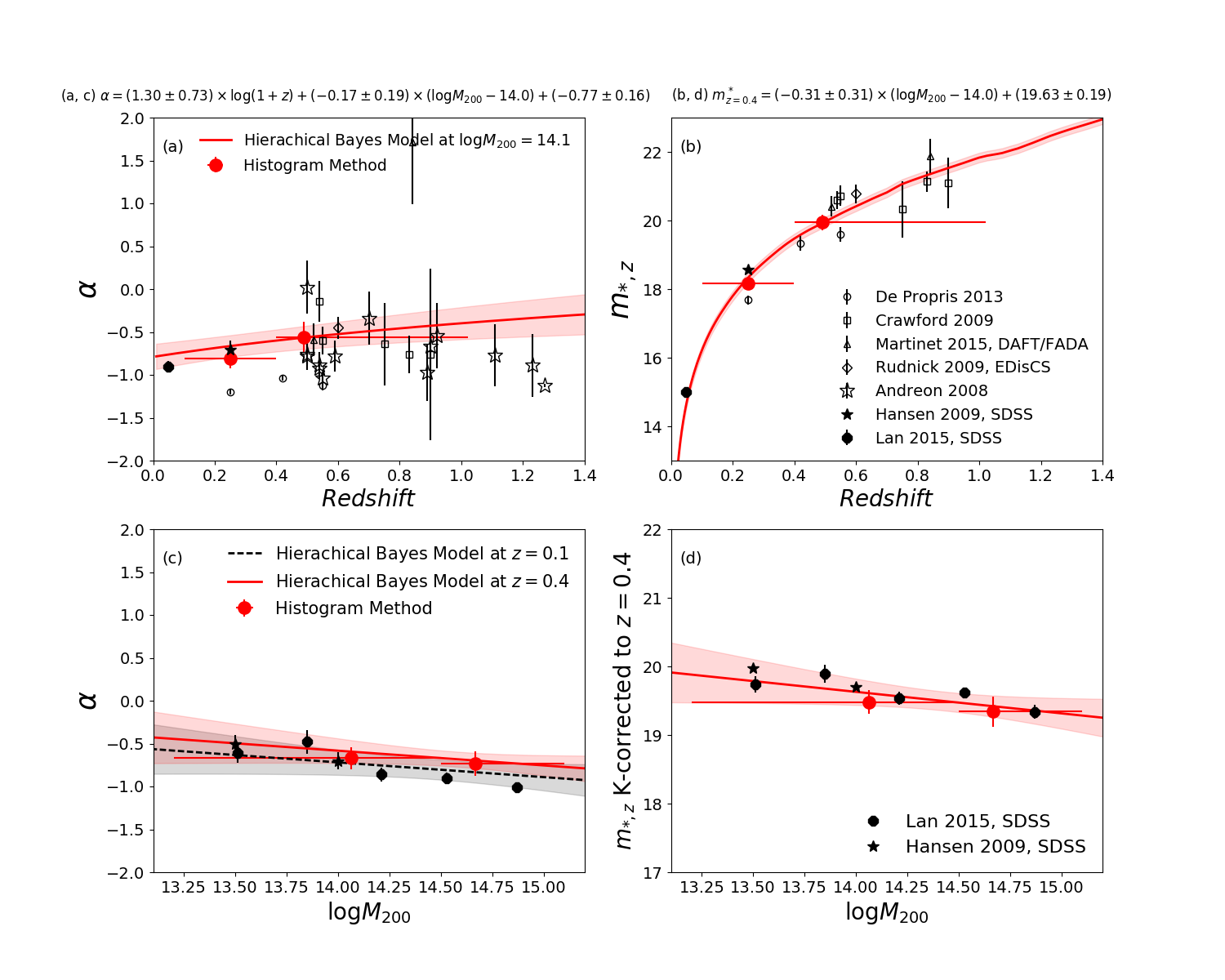}
\caption{ (Panels a and b) Redshift evolution of the faint end slope, $\alpha$, and the characteristic magnitude, $m^{\ast}$ (assuming passive redshift evolution of a simple stellar population from Bruzual \& Charlot (2003) 
 with a single star burst of metallicity $Z=0.008$ at $z=3.0$). (Panels c and d)  Mass dependence of the faint end slope, $\alpha$, and the characteristic magnitude, $m^{\ast}$ (assuming passive redshift evolution). Solid red lines and shades indicate results derived with the hierarchical Bayesian method (Section~\ref{sec: bayes}). Solid red circles indicate results derived with the alternative histogram method (Section~\ref{sec:histogram}). Literature reports of the $\alpha$ and $m^{\ast}$ parameters are over-plotted.}
\label{fig:redshift}
\end{figure*}

\subsection{Results from Hierarchical Bayesian Modeling}\label{sec:Bayes_res}

The Hierarchical Bayes model (Section \ref{sec:dependence}) simultaneously constrains the redshift evolution and mass dependence of $\alpha$ and $m_{*}$ anchored at redshift 0.4:
\begin{equation}
\begin{split}
\alpha & = A_\alpha \mathrm{log}(1+z) +B_\alpha (\mathrm{log} M_{200}-14) +C_\alpha \\
m^*_{z=0.4} & = B_m  (\mathrm{log} M_{200}-14) +C_m.
\end{split}
\end{equation}

The $m_{*}$ at other redshifts  are derived through evolving a passive redshift evolution model described in Section\ref{sec:dependence}.
\begin{table}
\begin{center}
\caption{Prior and Posterior Distributions of the parameters (see Equations~\ref{eq:z_mass_dependence},\ref{eq:phi_mass_dependence} and \ref{eq:phi_mass_dependence}) in the Hierarchical Bayes Model\label{tbl:alpha_mstar}}
\begin{tabular}{ccccccccc}
\hline
   & Prior & Posterior \\

$A_\alpha$ & [-5, 10] & $1.30\pm 0.70$ \\

$B_\alpha$ & [-4, 4] &  $-0.17\pm 0.19$ \\

$C_\alpha$ & [-2, 2] &  $-0.77\pm 0.16$ \\

$B_m$ & [-10, 10] & $-0.31\pm 0.31$ & \\

$C_m$  & $\mathcal{N}(-22.13, 1.0)$ & $-22.19\pm 0.19$  \\
  at $z=0.4$   &   $\mathcal{N}(19.69, 1.0) $         &$19.63\pm0.19$  \\

$B_\phi$ & [-5, 5] & $0.73\pm 0.13$ \\

$C_\phi$ & [-10, 10] & $0.85\pm 0.08$ \\
\hline

\end{tabular}
\end{center}
\end{table}

For each cluster, we only make use of the [$m^*$-2, $m^*$+2] magnitude range. Galaxy members of the analyzed clusters are complete within this range by selection (see details in Section~\ref{sec:cluster_sample}). The constraints of the $\alpha$ and $m^*_{z}$ relations are listed in Table~\ref{tbl:alpha_mstar}. The model posterior distributions are Gaussian-like according to visual checks. In Figure~\ref{fig:redshift} , we plot the $\alpha$ and $m^*_{z}$ relations as well as their uncertainties. 
For comparison, we show constraints from the alternative histogram approach (discussed in the following section).

The RSLF faint end slope, $\alpha$, displays a weak evidence of redshift evolution. The $A_\alpha$ parameter that controls the redshift evolution effect deviates from 0 at a significance level of $1.9\sigma$. For clusters of $\mathrm{log}M_{200}=14.1$ (median mass of the cluster sample),  $\alpha$ is constrained to be $-0.69 \pm 0.13$ at $z=0.2$, rising to $ - 0.52 \pm 0.14$ at $z=0.6$.
The mass dependence of $\alpha$ is ambiguous. The $B_\alpha$ parameter that controls this feature deviates from 0 by $ 0.9\sigma$. The effect is likely degenerate with the mass dependence of $m^*$. When removing $m^*$ mass dependence from the method (setting $B_m$ to be 0), $B_\alpha$ is consistent with 0.

We assume passive evolution to the RSLF characteristic magnitude $m^*_{z}$. We do not notice deviations of $m^*$ from the assumption (the $m^*$ results in redshift and mass bins agree with the model). Although the method models $m^*$ as mass-dependent, the effect appears to be insignificant ($B_m$ deviates from 0 by $1.0\sigma$).

\begin{figure}
\includegraphics[width=0.5\textwidth]{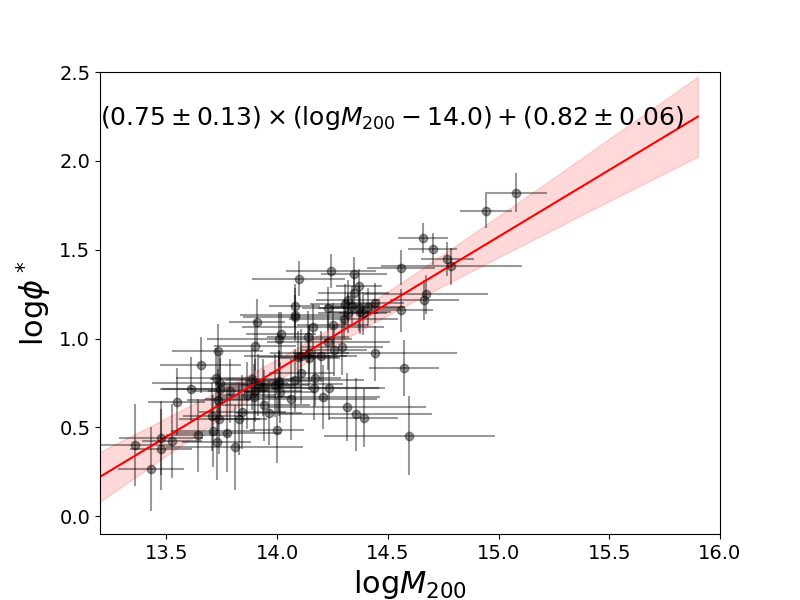}
\caption{Constraints of the RSLF amplitudes for individual clusters (black points). We model the RSLF amplitudes as mass dependent in the hierarchical Bayesian method (Section\ref{sec:dependence}). The solid line and shade show the constrained linear relation between $\mathrm{log}\phi^*$ and $ \mathrm{log}M_{200}$ as well as the $1\sigma$ uncertainty (intrinsic scatter of the relations is not constrained and hence not included in the uncertainty estimation).}
\label{fig:mass2}
\end{figure}

The hierarchical Bayesian method also constrains the RSLF amplitudes, $\phi^*$, and the relations between  $\phi^*$ and $\mathrm{log} M_{200}$.  $\phi^*$ scales with the total number of cluster galaxies. Our result shows a strong correlation between $\phi^*$ and the cluster mass (Figure~\ref{fig:mass2}). 

\subsection{Results in Redshift/Mass Bins}
\label{sec:res_z_mass}

\begin{table}
\begin{center}
\caption{Fitted Schechter Function parameters in redshift/mass bins} \label{tbl:bin_result}
\begin{tabular}{ccccccccc}
\hline
Cluster Selection
   & $\alpha$ &$m^*$& \\

$0.1\le z<0.4$ & $-0.80\pm 0.12$ & $18.17\pm0.18$ \\
64 clusters & & K-corrected to $z=0.25$ \\
 \\
$0.4\le z<1.05$ & $-0.55\pm 0.18$ & $19.96\pm0.23$ \\
27 clusters & & K-corrected to $z=0.49$\\
 \\
$13.2\le \mathrm{log} M_{200}<14.4$ &  $-0.67\pm0.12$ & $19.48\pm0.17$\\
77 clusters & & K-corrected to $z=0.4$\\
 \\
$14.4\le \mathrm{log} M_{200}<15.1$ & $-0.73\pm0.14$ & $19.34\pm0.22$\\
14 clusters & & K-corrected to $z=0.4$\\
\hline
\end{tabular}
\end{center}
\end{table}

We divide the clusters into two redshift bins: $0.1\le z<0.4$ and $0.4\le z<1.05$ and apply the alternative histogram method (Section~\ref{sec:histogram})\footnote{The reshift/mass cuts of the histogram samples are chosen by judgement to enlarge the redshift/mass differences of the subsets.}. The median cluster masses in each of the bins are $10^{14.1} \mathrm{M_\odot}$ and $10^{14.16}\mathrm{M_\odot}$ respectively.  The fitted parameters are listed in Table~\ref{tbl:bin_result}. Results are also shown in Figure~\ref{fig:hist} and~\ref{fig:redshift}. Again, the RSLF faint end slope, $\alpha$, displays a hint of redshift evolution. The measurements in two redshift bins differ by $\sim 1.2 \sigma$.
  
We divide the clusters into two mass bins: $13.2\le \mathrm{log} M_{200}<14.4$, $14.4\le \mathrm{log} M_{200}<15.1$ and apply the alternative histogram method. The median cluster redshifts in each of the bins are 0.35 and 0.34 respectively. To reduce uncertainties from band-shifting, we K-correct the RSLFs to $z=0.4$ (based on the red sequence model in Section~\ref{sec:select_red}). Results are presented in Table~\ref{tbl:bin_result}, Figure~\ref{fig:hist_mass} and Figure~\ref{fig:redshift}. No mass dependence of either $\alpha$ or $m^*$ is noted.
 
As shown in Figure~\ref{fig:redshift}, the results in cluster redshift/mass bins agree with the extrapolations from the hierarchical Bayesian model (Section~\ref{sec:Bayes_res}) within 1$\sigma$.
 
\subsection{Comparison to Literature}
\label{sec:literature}
In Figure~\ref{fig:redshift}, we over-plot literature measurements of the RSLF $\alpha$ and $m^*$ parameters. In the comparison datasets, \citet{2008MNRAS.386.1045A}, \citet{2009ApJ...700.1559R}, \citet{2009ApJ...690.1158C}, \citet{2013MNRAS.434.3469D}, and \citet{2015A&A...575A.116M} utilize smaller (N$_{clus}$ = 5-40) samples with individually measured LFs. For these, we compare to their stacked analyses when available since stacking reduce intrinsic cluster-to-cluster variations, something we achieve naturally in our Bayesian hierarchical model. We note that our Bayesian analysis utilizes a likelihood that is continuous in redshift, negating the need to stack our clusters in redshift bins (see Section \ref{sec: bayes}). We also include two low redshift constraints from stacked RSLFs on large cluster samples from the SDSS \citep{2009ApJ...699.1333H, 2015arXiv150907870L}. We do not compare to individual cluster RSLFs from the literature, since we do not have any known expectations on the cluster-to-cluster scatter in individual systems.

At low redshift, RSLF analyses based on SDSS data are available from \citet[$z\sim0.25$]{2009ApJ...699.1333H} and \citet[$z<0.05$]{2015arXiv150907870L}. The SDSS faint end slope measurements \citep{{2009ApJ...699.1333H}} appear to be  consistent with our results.  The SDSS characteristic magnitudes appears to be slightly fainter than the values constrained in this paper, but still consistent within this paper's 1$\sigma$ uncertainties ($M^*_z$ at redshift 0.4 is $-22.0$ from Lan et al. or $-22.13$ from Hansen et al., comparing to $-22.19\pm0.19$ in this paper). Note that the SDSS results are derived with $r$ \citep[$z<0.05$]{2015arXiv150907870L} or $i$ \citep[$z<0.05$]{2015arXiv150907870L} band data and we assume a red sequence model in Section~\ref{sec:select_red} when comparing the characteristic magnitudes.

In terms of the parameter mass dependence, the $\alpha$ and $m^*$ measurements from \citet[$z<0.05$]{2015arXiv150907870L} in different cluster mass ranges match well with our constraints. In \cite{2009ApJ...699.1333H}, the mass dependence results for cluster RS galaxies are not explicitly listed, but there is a trend of $\alpha$ steepening in the mass range of [$10^{13}\mathrm{M}_\odot$, $10^{14}\mathrm{M}_\odot$], and then stabilizing beyond $10^{14}\mathrm{M}_\odot$.  The quantity $m^*$ displays a trend of brightening in the mass range of [$10^{13}\mathrm{M}_\odot$, $5\times10^{14}\mathrm{M}_\odot$], and then stabilizing beyond $5\times10^{14}\mathrm{M}_\odot$. These measurements qualitatively agree to our result. 

At intermediate to high redshift, measurements of RSLF are still scarce. Sample sizes used in previous works are much smaller than those in SDSS-based studies. Any mass dependent effect of $\alpha$ would make it difficult to make a direct comparison in Figure~\ref{fig:redshift}.  \citet{2008MNRAS.386.1045A} measures individual LFs for 16 clusters at $z > 0.5$, which we include on Figure \ref{fig:redshift} a,b. We caution that comparing our results to these data is problematic for two reasons. First, the \citet{2008MNRAS.386.1045A} clusters have RSLFs measured using galaxy data extracted from a fixed observed angle that corresponds to a smaller projected radii than we use. We utilize a fixed co-moving radius, thus minimizing any radial evolution that might be present. Second, our Bayesian RSLF technique smooths out cluster-to-cluster scatter, similar to stacking. On the other hand, interpreting individual cluster RSLFs requires that the specific (and small) sample be representative of the mean population.
A closer comparison to our dataset is to \citet{2015A&A...575A.116M}. They create two stacked clusters, one based on about a dozen clusters at $\langle z \rangle = 0.5$ and one based on 3 or 4 clusters at $\langle z \rangle = 0.84$. They use a fixed 1Mpc radius for their galaxy extraction. We find good agreement, although their error bars are much larger.

Our sample makes a significant contribution to the observed evolution of the RSLF through its quality, size, redshift coverage, and mass range. Compared to current RSLF analyses, our DES/XCS sample is one of the very few that we can expect cluster-to-cluster variations to be minimized over a large redshift range of $0.2 \le z \le 1$. We are able to constrain the RSLF over the entire redfshift range without combining disparate results at different redshifts. With a single dataset, we eliminate issues that could be created by heterogeneity from instrumentation, photometry, statistical techniques, etc. At the same time, by having X-ray inferred cluster masses, we are able to account for covariance in slope evolution between redshift and cluster mass.
 
\section{Discussion and Summary}
\label{sec:discussion}

This paper constrains the evolution of the red sequence luminosity function (RSLF). Typically, the cluster luminosity function has been studied using clusters with well-sampled data (i.e., deep observations) or through stacking/averaging clusters \citep{2008ApJ...676..248Y, 2009ApJ...699.1333H,2008MNRAS.386.1045A, 2009ApJ...700.1559R, 2009ApJ...690.1158C, 2013MNRAS.434.3469D, 2015A&A...575A.116M}. While our DES observations are fairly deep, we utilize stringent completeness limits in order to avoid any complications with modeling the faint end slope. This means that the data on any individual cluster may not be good enough to measure the RSLF with traditional statistical techniques, especially at high z. At the same time, stacking has its own concerns. \cite{2009ApJ...690.1158C} discussed possible caveats when  interpreting stacked luminosity functions. For instance, cluster luminosity function stacks could be biased by clusters that have brighter $m_*$ or more negative $\alpha$. Thus, the interpretation of the stacked $m_*$ and $\alpha$ is complicated. 

In this paper, we bridge the gap between the above two standard RSLF techniques by employing a hierarchical Bayesian model. This models allows us to use the sparse and noisy data from the individual clusters, while at the same time incorporating prior information (e.g., from the X-ray inferred cluster masses). We develop a model which allows the faint-end slope of the RSLF (parametrized as $\alpha$) to be a function of the log of both the cluster mass and redshift. The model also allows $m^{*}$ and the overall RSLF amplitude $\phi^{*}$ to vary linearly with the log of the cluster cluster mass. 

Using this hierarchical Bayesian model on a sample of 94 X-ray select clusters to a $z= 1.05$, we find weak (1.9$\sigma$) evidence of redshift evolution for the RSLF faint end slope. Redshift evolution in the shape of the RSLF could indicate a rising abundance of faint RS galaxies over time. The result is consistent with a non-evolving fraction of cluster red galaxies to $z\sim 1$ in clusters. For consistency, we bin the clusters according to redshift and mass and stack the red sequence galaxies to increase the signal-to-noise of the RSLF. The stacked RSLF parameters are consistent with the Bayesian results. Our work represents one of the largest RSLF studies to date that goes to redshift $\sim$ 1.0.

\begin{figure}
\includegraphics[width=0.5\textwidth]{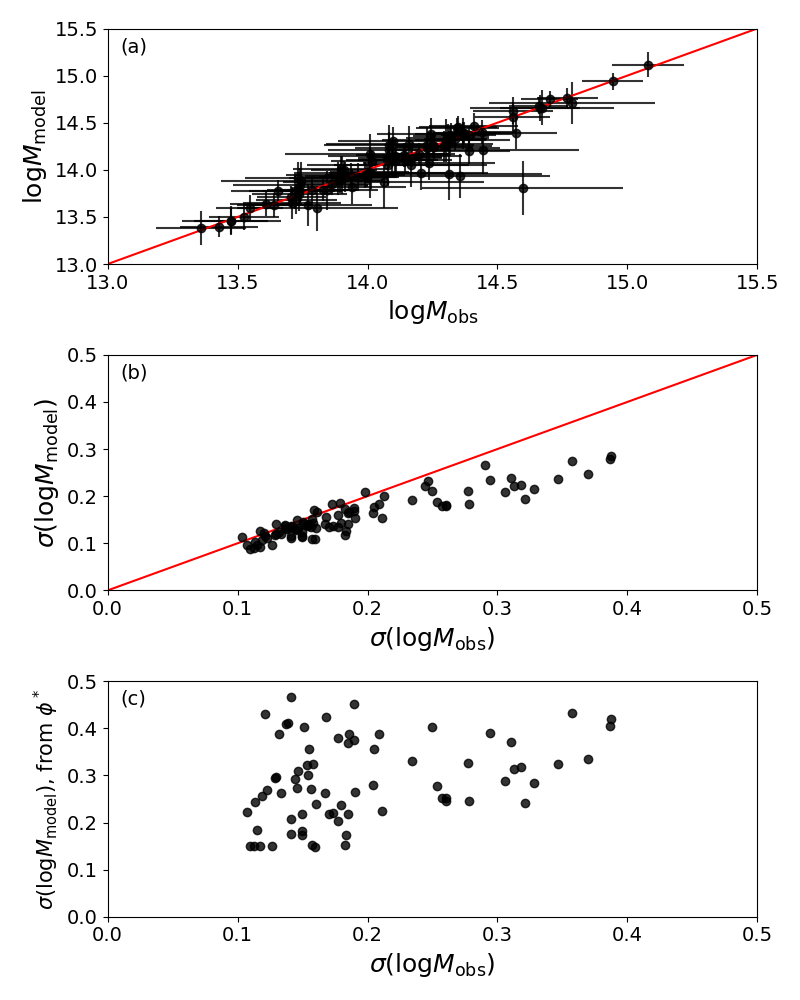}
\caption{ In the hierarchical Bayesian method, we constrain cluster masses using X-ray temperature-inferred measurements as priors. (Panel a)  The posterior estimations of cluster masses, $\mathrm{log}M_\mathrm{model}$, agree with the priors $\mathrm{log}M_\mathrm{obs}$. (Panel b)  The assumption in the hierarchical Bayes model that  cluster masses scale with RSLF amplitudes, $\phi^*$, helps improving the accuracy of cluster mass estimations. The posterior uncertainties of the mass estimations, $\sigma(\mathrm{log}M_\mathrm{model})$, appear to be decreased, especially when the prior uncertainties, $\sigma(\mathrm{log}M_\mathrm{obs})$, are higher than 0.2 dex. (Panel c) Based on the improved  estimation on the values of cluster mass ($\mathrm{log}M_\mathrm{model}$) we estimate the uncertainties of inferring cluster masses from  $\phi^*$ only, which range from 0.2 to 0.4 dex (see details in Section~\ref{sec:Bayes_res}). }
\label{fig:mass_err}
\end{figure}

A particularly interesting by-product of this study is that our model allows us to improve the cluster mass estimation. This is because our Bayesian model allows cluster mass estimation, $\mathrm{log}M_\mathrm{model}$,  to deviate from its prior values inferred from X-ray measurements ($\mathrm{log}M_\mathrm{obs}$) by considering the correlation between $\phi^*$ and cluster mass. While the posterior values of cluster mass agree to its prior values ($\mathrm{log}M_\mathrm{model}$ compared to $\mathrm{log}M_\mathrm{obs}$ in the top panel of Figure~\ref{fig:mass_err}), the precision of the mass estimations have been improved as indicated by their smaller posterior uncertainties ($\sigma(\mathrm{log}M_\mathrm{model})$ compared to $\sigma(\mathrm{log}M_\mathrm{obs})$ in the middle panel of Figure~\ref{fig:mass_err}). The improvements are especially noticeable when the mass prior uncertainties -- $\sigma(\mathrm{log}M_\mathrm{obs})$, which include both the intrinsic scatter of the X-ray observable-mass scaling relations and statistical uncertainties of the observable -- is higher than 0.3 dex. 

Based on the improved estimation on the values of $\mathrm{log}M_\mathrm{model}$, and assuming $\phi^*$ and X-ray measurements contribute independent Gaussian-like intrinsic and measurement uncertainties to $\mathrm{log}M_\mathrm{model}$, 
\begin{equation}
 \frac{1}{\sigma^2(\mathrm{log}M_\mathrm{obs})} + \frac{1}{\sigma^2(\mathrm{log}M_\mathrm{model})\text{ from }\phi^*} = \frac{1}{ \sigma^2(\mathrm{log}M_\mathrm{model}) },
\end{equation}
we estimate the uncertainties of inferring cluster mass from only  $\phi^*$ as:
\begin{equation}
 \sigma(\mathrm{log}M_\mathrm{model})\text{ from }\phi^* = \frac{ \sigma(\mathrm{log}M_\mathrm{model}) }{\sqrt{1.0-\frac{\sigma^2(\mathrm{log}M_\mathrm{model})}{\sigma^2(\mathrm{log}M_\mathrm{obs}) }}}.
\end{equation}
These estimations are shown in the bottom panel of Figure~\ref{fig:mass_err}, which range from 0.2 to 0.4, with an average of 0.34. Because estimating cluster mass from $\phi^*$ is physically driven by the cluster galaxy over densities and thus sensitive to the presence of foreground and background galaxies, these mass uncertainties tends to be much larger than the X-ray temperature derived mass uncertainties. Comparatively, optical mass proxies derived from the numbers of cluster galaxies have intrinsic mass scatters between 0.2 to 0.5 dex \citep{2009ApJ...699..768R, 2015MNRAS.454.2305S}. This analysis demonstrates the potential of $\phi^*$ as a cluster mass proxy.

Since the redshift evolution of the RSLF is only insignificantly detected at a significance level of $1.9\sigma$, it is worthwhile to apply the analysis to a larger cluster sample. We expect the XCS to find over 1000 clusters within the DES final data release. We may also utilize new and large optical cluster catalogs such as RedMaPPer. However, optically characterized clusters will add new challenges from the covariance between the richness-inferred cluster masses and the red-sequence luminosity functions.  An evolving abundance of faint RS galaxies will also introduce a redshift evolution component into the cluster mass-richness scaling relation. Assuming the $\alpha$ evolution reported in this paper, we expect the number of RS galaxies above $m^*+2~\mathrm{mag}$ to decrease by $\sim 20\%$ from $z=0$  to $z=1.0$. Using the parameterization of cluster mass-richness scaling relation in \cite{2016arXiv161006890M}, we expect the mass-to-richness ratio to change with redshift as $(1+z)^{0.26}$ (constrained as $(1+z)^{0.18\pm0.75(\mathrm{stat})\pm0.24(\mathrm{sys})}$ in the fore-mentioned weak lensing study). Of course there could be additional effects on the mass-richness relation if there is redshift evolution in $m^*$ and $\phi^*$ or if the mass dependence of the RSLF is not properly accounted for. 

Regardless, we expect to increase the X-ray cluster sample size by at least a factor of 10 by the end of DES, covering a similar redshift range with this analysis. Using catalog-level simulations of RSLF similar to the ones observed here, we expect to increase our sensitivity on the evolution of $\alpha$ by a factor of three.

If there is redshift evolution in the faint-end slope of the red sequence galaxies, we can explain it through the formation times and growth histories of galaxies. For instance, bright and faint cluster red sequence galaxies may have different formation times. It is possible that fainter galaxies are quenched during, rather than before, the cluster infall process. Hence the fraction of faint red sequence galaxies gradually increase with time. Astrophysical processes that slowly shut down galaxy star formation activities, e.g., strangulation (sometimes called starvation) \citep{1980ApJ...237..692L, 2000MNRAS.318..703B, 2000ApJ...540..113B, 2015Natur.521..192P} and hence gradually increase the fraction of faint red sequence galaxies, will be preferred over more rapid processes such as ram pressure stripping \citep{1972ApJ...176....1G, 2000Sci...288.1617Q}. Combining the observational constraints on the evolution of the faint-end slope together with the cluster accretion history in simulations should help us place good constraints on the formation and transition times of cluster red sequence galaxies \citep{2009MNRAS.400..937M}.

\label{sec:summary}

%@YyZfromKR: I'd have split the sentence at the , in "size of the sample, when the significance"

In summary, we constrain the relation between RSLF amplitudes and cluster masses, and the correlation improves the estimation of cluster masses. We find a hint that the Schechter function faint-end slope becomes less negative for clusters at higher redshifts, indicating a rising abundance of faint red sequence galaxies with time. The redshift evolution of RSLF parameters may also impact the accuracy of optical cluster cosmology analyses. These results are acquired with a hierarchical Bayesian method, which has the advantage of  disentangling simultaneous RSLF dependence on cluster mass and redshift despite the small size of the sample. The significance of the results would have been easily underlooked by a stacking method, which is also tested in this paper.

\section*{Acknowledgements}
C. Miller as well as Y. Zhang acknowledges support from Department of Energy grant DE-SC0013520. Y. Zhang thanks Alex Drlica-Wagner, Gary Bernstein and Chris Davis for careful reading of the draft.
We use DES Science Verification in this paper. We are grateful for the extraordinary contributions of our CTIO colleagues and the DECam Construction, Commissioning and Science Verification
teams in achieving the excellent instrument and telescope conditions that have made this work possible.  The success of this project also 
relies critically on the expertise and dedication of the DES Data Management group.

Funding for the DES Projects has been provided by the U.S. Department of Energy, the U.S. National Science Foundation, the Ministry of Science and Education of Spain, 
the Science and Technology Facilities Council of the United Kingdom, the Higher Education Funding Council for England, the National Center for Supercomputing 
Applications at the University of Illinois at Urbana-Champaign, the Kavli Institute of Cosmological Physics at the University of Chicago, 
the Center for Cosmology and Astro-Particle Physics at the Ohio State University,
the Mitchell Institute for Fundamental Physics and Astronomy at Texas A\&M University, Financiadora de Estudos e Projetos, 
Funda{\c c}{\~a}o Carlos Chagas Filho de Amparo {\`a} Pesquisa do Estado do Rio de Janeiro, Conselho Nacional de Desenvolvimento Cient{\'i}fico e Tecnol{\'o}gico and 
the Minist{\'e}rio da Ci{\^e}ncia, Tecnologia e Inova{\c c}{\~a}o, the Deutsche Forschungsgemeinschaft and the Collaborating Institutions in the Dark Energy Survey. 

The Collaborating Institutions are Argonne National Laboratory, the University of California at Santa Cruz, the University of Cambridge, Centro de Investigaciones Energ{\'e}ticas, 
Medioambientales y Tecnol{\'o}gicas-Madrid, the University of Chicago, University College London, the DES-Brazil Consortium, the University of Edinburgh, 
the Eidgen{\"o}ssische Technische Hochschule (ETH) Z{\"u}rich, 
Fermi National Accelerator Laboratory, the University of Illinois at Urbana-Champaign, the Institut de Ci{\`e}ncies de l'Espai (IEEC/CSIC), 
the Institut de F{\'i}sica d'Altes Energies, Lawrence Berkeley National Laboratory, the Ludwig-Maximilians Universit{\"a}t M{\"u}nchen and the associated Excellence Cluster Universe, 
the University of Michigan, the National Optical Astronomy Observatory, the University of Nottingham, The Ohio State University, the University of Pennsylvania, the University of Portsmouth, 
SLAC National Accelerator Laboratory, Stanford University, the University of Sussex, Texas A\&M University, and the OzDES Membership Consortium.

Based in part on observations at Cerro Tololo Inter-American Observatory, National Optical Astronomy Observatory, which is operated by the Association of 
Universities for Research in Astronomy (AURA) under a cooperative agreement with the National Science Foundation.

The DES data management system is supported by the National Science Foundation under Grant Numbers AST-1138766 and AST-1536171.
The DES participants from Spanish institutions are partially supported by MINECO under grants AYA2015-71825, ESP2015-88861, FPA2015-68048, SEV-2012-0234, SEV-2016-0597, and MDM-2015-0509, 
some of which include ERDF funds from the European Union. IFAE is partially funded by the CERCA program of the Generalitat de Catalunya.
Research leading to these results has received funding from the European Research
Council under the European Union's Seventh Framework Program (FP7/2007-2013) including ERC grant agreements 240672, 291329, and 306478.
We  acknowledge support from the Australian Research Council Centre of Excellence for All-sky Astrophysics (CAASTRO), through project number CE110001020.

This manuscript has been authored by Fermi Research Alliance, LLC under Contract No. DE-AC02-07CH11359 with the U.S. Department of Energy, Office of Science, Office of High Energy Physics. The United States Government retains and the publisher, by accepting the article for publication, acknowledges that the United States Government retains a non-exclusive, paid-up, irrevocable, world-wide license to publish or reproduce the published form of this manuscript, or allow others to do so, for United States Government purposes.

\section*{Affiliations}
{\small
%\begin{enumerate}[label=$^{\arabic*}\,$, leftmargin=*, align=left]
\affil{$^{1}$ Fermi National Accelerator Laboratory, P. O. Box 500, Batavia, IL 60510, USA}
\affil{$^{2}$ Department of Astronomy, University of Michigan, Ann Arbor, MI 48109, USA}
\affil{$^{3}$ Department of Physics, University of Michigan, Ann Arbor, MI 48109, USA}
\affil{$^{4}$ Department of Physics and Astronomy, Pevensey Building, University of Sussex, Brighton, BN1 9QH, UK}
\affil{$^{5}$ Kavli Institute for Particle Astrophysics \& Cosmology, P. O. Box 2450, Stanford University, Stanford, CA 94305, USA}
\affil{$^{6}$ SLAC National Accelerator Laboratory, Menlo Park, CA 94025, USA}
\affil{$^{7}$ Faculty of Physics, Ludwig-Maximilians-Universit\"at, Scheinerstr. 1, 81679 Munich, Germany}
\affil{$^{8}$ Excellence Cluster Universe, Boltzmannstr.\ 2, 85748 Garching, Germany}
\affil{$^{9}$  Taejon Christian International School, Yuseong, Daejeon, 34035, South Korea}
\affil{$^{10}$ Institute of Cosmology \& Gravitation, University of Portsmouth, Portsmouth, PO1 3FX, UK}
\affil{$^{11}$ Jodrell Bank Center for Astrophysics, School of Physics and Astronomy, University of Manchester, Oxford Road, Manchester, M13 9PL, UK}
\affil{$^{12}$ Astrophysics Research Institute, Liverpool John Moores University, IC2, Liverpool Science Park, 146 Brownlow Hill, Liverpool L3 5RF, UK}
\affil{$^{13}$ University of Nottingham, School of Physics and Astronomy, Nottingham NG7 2RD, UK}
\affil{$^{14}$ Astrophysics and Cosmology Research Unit, School of Mathematics, Statistics and Computer Science, University of KwaZulu\-Natal, Westville Campus, Durban 4000, South Africa}
\affil{$^{15}$ Universit\"ats-Sternwarte, Fakult\"at f\"ur Physik, Ludwig-Maximilians Universit\"at M\"unchen, Scheinerstr. 1, 81679 M\"unchen, Germany}
\affil{$^{16}$ Jodrell Bank Centre for Astrophysics, School of Physics and Astronomy, The University of Manchester, Manchester M13 9PL, UK}
\affil{$^{17}$ Institute for Astronomy, University of Edinburgh, Edinburgh EH9 3HJ, UK}
\affil{$^{18}$ Institute for Astronomy, University of Edinburgh, Royal Observatory, Blackford Hill, Edinburgh EH9 3HJ, United Kingdom}
\affil{$^{19}$ George P. and Cynthia Woods Mitchell Institute for Fundamental Physics and Astronomy, and Department of Physics and Astronomy, Texas A\&M University, College Station, TX 77843,  USA}
\affil{$^{20}$ Department of Physics and Astronomy, Uppsala University, Box 516, SE-751 20 Uppsala, Sweden}
\affil{$^{21}$ Physics Department, Lancaster University, Lancaster LA1 4YB, UK}
\affil{$^{22}$ Instituto de Astrof\'{\i}sica e Ci\^{e}ncias do Espa\c{c}o, Universidade do Porto, CAUP, Rua das Estrelas, 4150-762 Porto, Portugal}
\affil{$^{23}$ Departamento de F\'{\i}sica e Astronomia, Faculdade de Ci\^{e}ncias, Universidade do Porto, Rua do Campo Alegre 687, 4169-007 Porto, Portugal}
\affil{$^{24}$ Department of Physics, Stanford University, 382 Via Pueblo Mall, Stanford, CA 94305, USA}
\affil{$^{25}$ Cerro Tololo Inter-American Observatory, National Optical Astronomy Observatory, Casilla 603, La Serena, Chile}
\affil{$^{26}$ Department of Physics \& Astronomy, University College London, Gower Street, London, WC1E 6BT, UK}
\affil{$^{27}$ Department of Physics and Electronics, Rhodes University, PO Box 94, Grahamstown, 6140, South Africa}
\affil{$^{28}$ CNRS, UMR 7095, Institut d'Astrophysique de Paris, F-75014, Paris, France}
\affil{$^{29}$ Sorbonne Universit\'es, UPMC Univ Paris 06, UMR 7095, Institut d'Astrophysique de Paris, F-75014, Paris, France}
\affil{$^{30}$ Laborat\'orio Interinstitucional de e-Astronomia - LIneA, Rua Gal. Jos\'e Cristino 77, Rio de Janeiro, RJ - 20921-400, Brazil}
\affil{$^{31}$ Observat\'orio Nacional, Rua Gal. Jos\'e Cristino 77, Rio de Janeiro, RJ - 20921-400, Brazil}
\affil{$^{32}$ Department of Astronomy, University of Illinois, 1002 W. Green Street, Urbana, IL 61801, USA}
\affil{$^{33}$ National Center for Supercomputing Applications, 1205 West Clark St., Urbana, IL 61801, USA}
\affil{$^{34}$ Institut de F\'{\i}sica d'Altes Energies (IFAE), The Barcelona Institute of Science and Technology, Campus UAB, 08193 Bellaterra (Barcelona) Spain}
\affil{$^{35}$ Institute of Space Sciences, IEEC-CSIC, Campus UAB, Carrer de Can Magrans, s/n,  08193 Barcelona, Spain}
\affil{$^{36}$ Department of Physics and Astronomy, University of Pennsylvania, Philadelphia, PA 19104, USA}
\affil{$^{37}$ Department of Physics, California Institute of Technology, Pasadena, CA 91125, USA}
\affil{$^{38}$ Jet Propulsion Laboratory, California Institute of Technology, 4800 Oak Grove Dr., Pasadena, CA 91109, USA}
\affil{$^{39}$ Instituto de Fisica Teorica UAM/CSIC, Universidad Autonoma de Madrid, 28049 Madrid, Spain}
\affil{$^{40}$ Center for Cosmology and Astro-Particle Physics, The Ohio State University, Columbus, OH 43210, USA}
\affil{$^{41}$ Department of Physics, The Ohio State University, Columbus, OH 43210, USA}
\affil{$^{42}$ Astronomy Department, University of Washington, Box 351580, Seattle, WA 98195, USA}
\affil{$^{43}$ Santa Cruz Institute for Particle Physics, Santa Cruz, CA 95064, USA}
\affil{$^{44}$ Australian Astronomical Observatory, North Ryde, NSW 2113, Australia}
\affil{$^{45}$ Departamento de F\'isica Matem\'atica, Instituto de F\'isica, Universidade de S\~ao Paulo, CP 66318, S\~ao Paulo, SP, 05314-970, Brazil}
\affil{$^{46}$ Department of Astrophysical Sciences, Princeton University, Peyton Hall, Princeton, NJ 08544, USA}
\affil{$^{47}$ Instituci\'o Catalana de Recerca i Estudis Avan\c{c}ats, E-08010 Barcelona, Spain}
\affil{$^{48}$ Centro de Investigaciones Energ\'eticas, Medioambientales y Tecnol\'ogicas (CIEMAT), Madrid, Spain}
\affil{$^{49}$ School of Physics and Astronomy, University of Southampton,  Southampton, SO17 1BJ, UK}
\affil{$^{50}$ Instituto de F\'isica Gleb Wataghin, Universidade Estadual de Campinas, 13083-859, Campinas, SP, Brazil}
\affil{$^{51}$ Computer Science and Mathematics Division, Oak Ridge National Laboratory, Oak Ridge, TN 37831}
%\end{enumerate}
}

\appendix
\section{Completeness Function}
\label{sec:completeness}

\subsection{The Completeness Function Model}

The completeness function models the detection probability of objects in terms of their apparent magnitude.
In this paper, the completeness function is defined as the ratio between the numbers of observed and true objects at magnitude $m$.

We model the completeness function with a complementary error function \citep{2011ApJ...734....3Z} of three parameters:
\begin{equation}
p(m)=\lambda \frac{1}{2}\mathrm{erfc}(\frac{m-m_{50}}{\sqrt{2w}}).
\label{eq:compl_func}
\end{equation}
In the above equation, $m_{50}$ is the  50\% completeness magnitude, $w$ controls the steepness of the detection drop-out rate and $\lambda$ is the overall amplitude of the completeness function.
We further assume linear dependence  of $m_{50}$ and $w$ on the depth of the image, which is characterized by the 10 $\sigma$ limiting magnitude\footnote{Magnitude with $magerr\_auto=$ 0.108. For a flux measurement at a significance level of 10 $\sigma$, the corresponding magnitude uncertainty is $ 2.5/ \mathrm{ln} 10/10=0.108$.}. In this paper, we evaluate the $z$-band completeness function, which is related to  image depth in $z$.

\begin{figure*}
\begin{center}
\includegraphics[width=0.75\textwidth]{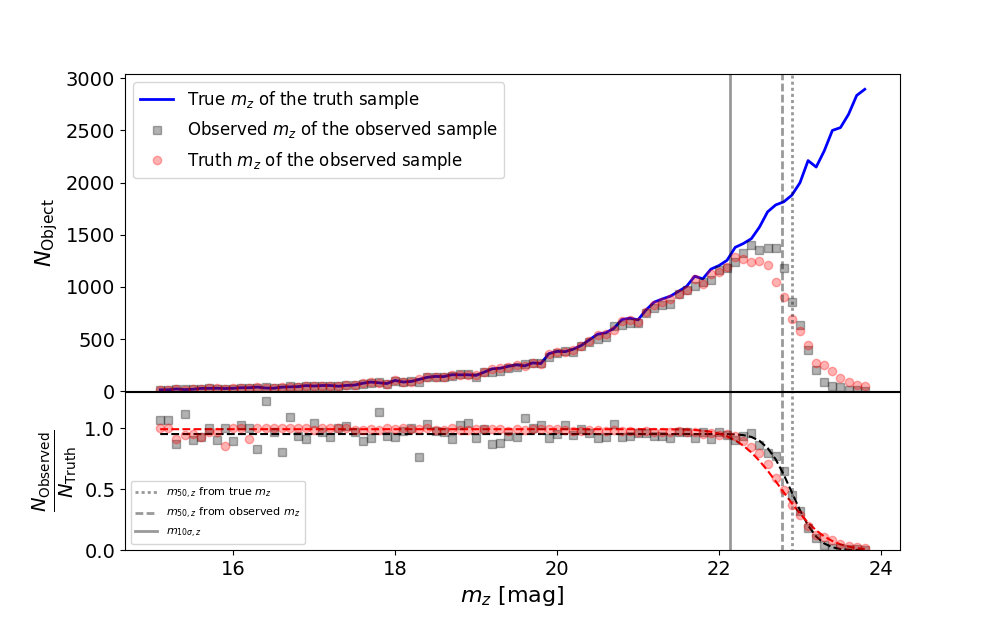}
\caption{This figure demonstrates our procedure for evaluating completeness function with the~\UFIG~simulation. We model the difference between the observed magnitude distribution (gray squares in the upper panel) of observed objects and the true magnitude distribution of all truth objects (solid blue line in the top panel). We model the ratio between the observed magnitude distribution and the truth magnitude distribution (gray squares in the lower panel) with a complementary error function (black dashed line). For comparison, we also show ratios between the truth magnitude distributions of the observed and the truth objects (red circles) and the complementary error function fitted model (red dashed line).} 
\label{fig:ufig_method}
\end{center}
\end{figure*}

\subsection{Relations between Model parameters and Image Depth}
\label{sec:cf_relations}

\begin{figure*}
\includegraphics[width=1.0\textwidth]{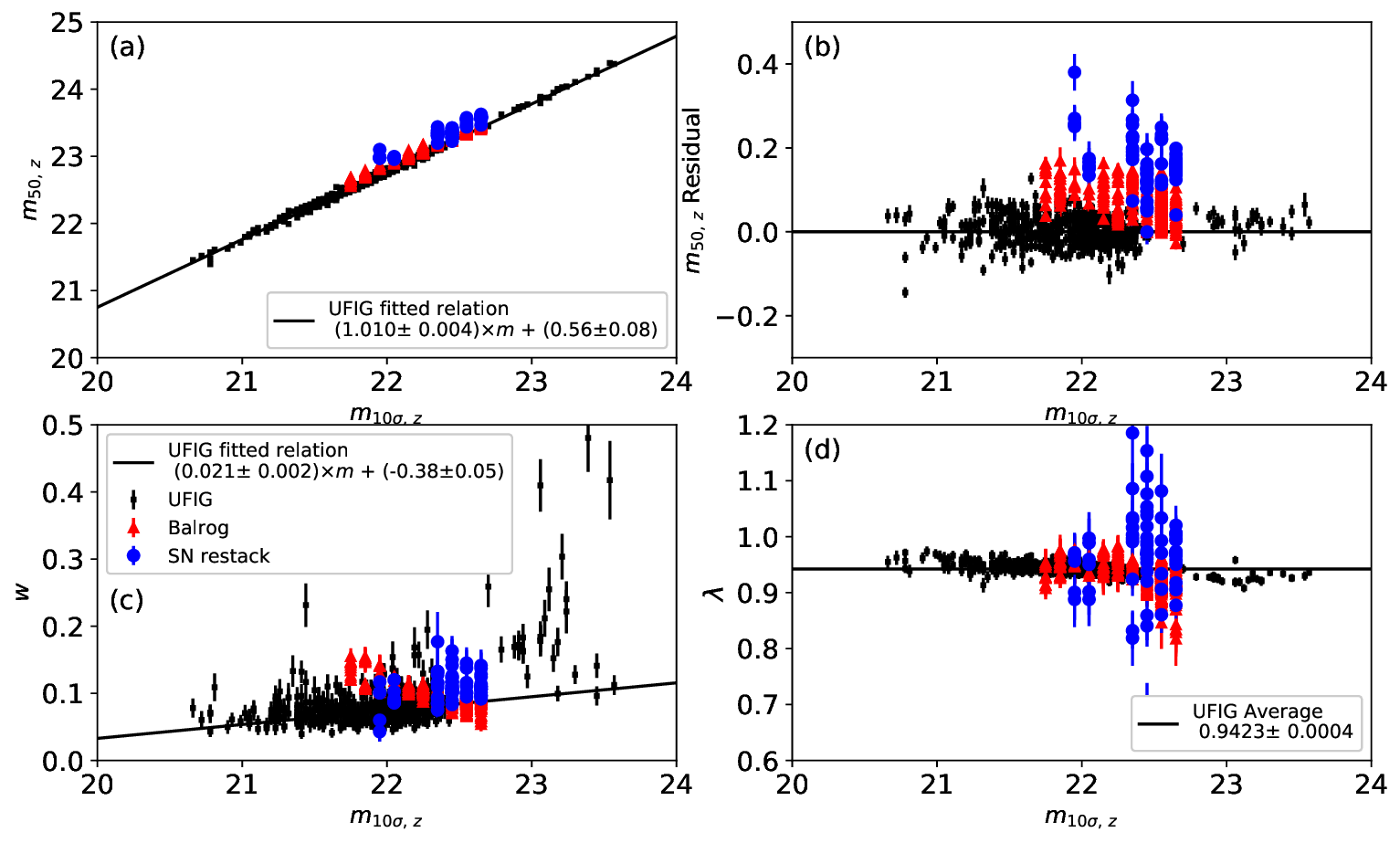}
\caption{This figure shows the relations between completeness function parameters and the image depth, characterized by the 10 $\sigma$ limiting magnitude. Panel (a) shows the dependences of $m_{50}$, the  50\% completeness magnitude, on image depth from the  \UFIG~(black points), \BALROG~(red triangles) simulations and the SN restack data (blue circles). Panel (b) shows the $m_{50}$ residuals of the three data sets from the \UFIG~ relation. The relation derived with the \UFIG~simulation generally agrees with the data from the \BALROG~simulation. The $m_{50}$ values evaluated from re-stacking deep supernovae data appear to be 0.1-0.2 mag deeper, but the differences can be explained by the Kron magnitude bias shown in Z16.  Panel (c) shows the dependences of $w$, the steepness of the detection drop-out rate, on image depth. We use the \UFIG~simulation relations for both $m_{50}$ and $w$ in this paper. We notice that the completeness function amplitudes from simulations appear to be lower than 1 as shown in panel (d), but it is mostly caused by the differences between observed and truth magnitudes (see a discussion in Section \ref{sec:cf_relations}).}
\label{fig:completeness_function}
\end{figure*}

The $m_{50}$ - $m_{10\sigma}$ and $w$ - $m_{10\sigma}$  relations are evaluated with simulated DES images and real data. The relations used in this paper are derived from the \UFIG~simulation \citep[][]{2013A&C.....1...23B, 2015ApJ...801...73C}\citep[also see][for an application]{2016ApJS..226...24L}, which is a sky simulation that is further based on an N-body dark matter simulation. The dark matter simulation is populated with galaxies from the Adding Density Determined GAlaxies to Lightcone Simulations (ADDGALS) algorithm \citep{2019arXiv190102401D}. 

We use the \UFIG~product that matches the footprint of the ``gold" sample in Section~\ref{sec:sva1_gold}. The simulation is divided into fields of 0.53 deg$^2$, with characteristic quantities like the image depth and seeing matching those of the DES-SV patches. SExtractor is applied to the simulated images with identical DES-SV settings. We select objects with $magerr\_auto < 0.218$ mag in $z$ (5$\sigma$ significance),  derive their observed magnitude distribution, and compare it to the truth magnitude distribution of all input truth objects (see illustration about the procedure in Figure~\ref{fig:ufig_method}). The ratio between the two is well described by Equation~\ref{eq:compl_func}. The derived $m_{50}$ and $w$ are tightly related to the depth of the image as shown in Figure ~\ref{fig:completeness_function}. 

We also perform the analysis with the \BALROG~simulation \citep{2015arXiv150708336S}, which inserts simulated objects into real DES-SV images. The results are similar.

 To further verify the derived relations,  we stack high quality images from the DES Supernovae survey  (with a total exposure time of $\sim$  1000 $s$) to mimic main survey depth. The $z$-band depth of the stacks ranges from 21.5 mag to 22.5 mag, comparing to $>24$ mag coadding all eligible exposures. We compare the object counts in this set of coadds and the full coadds to evaluate $m_{50}$ and $w$ (also shown in Figure ~\ref{fig:completeness_function}). 
 
 The $m_{50}$ appears to be 0.1 - 0.4 mag deeper than  the simulation relations. The effect is  consistent with the $mag\_auto$ biases shown in Z16. In this test, we compare to the observed Kron magnitudes rather than the "truth" magnitudes (which is not known) from the deeper stack. Z16 shows that the observed Kron magnitudes  are fainter by 0.1 to 0.4 mag comparing to the "truth" magnitudes  at < 24 mag.
 
Figure~\ref{fig:completeness_function} indicates that the amplitude of the complementary error function is lower than 1 in \UFIG  ~and \BALROG.  This is mostly caused by the same photometry measurement bias discussed above (another effect is the blending of truth objects, which causes incompleteness at a $<2$\% level). Objects are measured fainter by the Kron magnitude. Compared to the truth magnitude distribution, the observed magnitude distribution is systematically shifted  to the fainter side (see this effect in Figure~\ref{fig:ufig_method}). The result is that the observed magnitude distribution is always lower than the truth distribution, and the amplitude of the fitted completeness function is below 1. This shift and the resulting amplitudes of the completeness function are not of interest in this paper. We explicitly assumes the amplitudes of the completeness function to be 1.
 
\begin{figure}
\begin{center}
\includegraphics[width=0.5\textwidth]{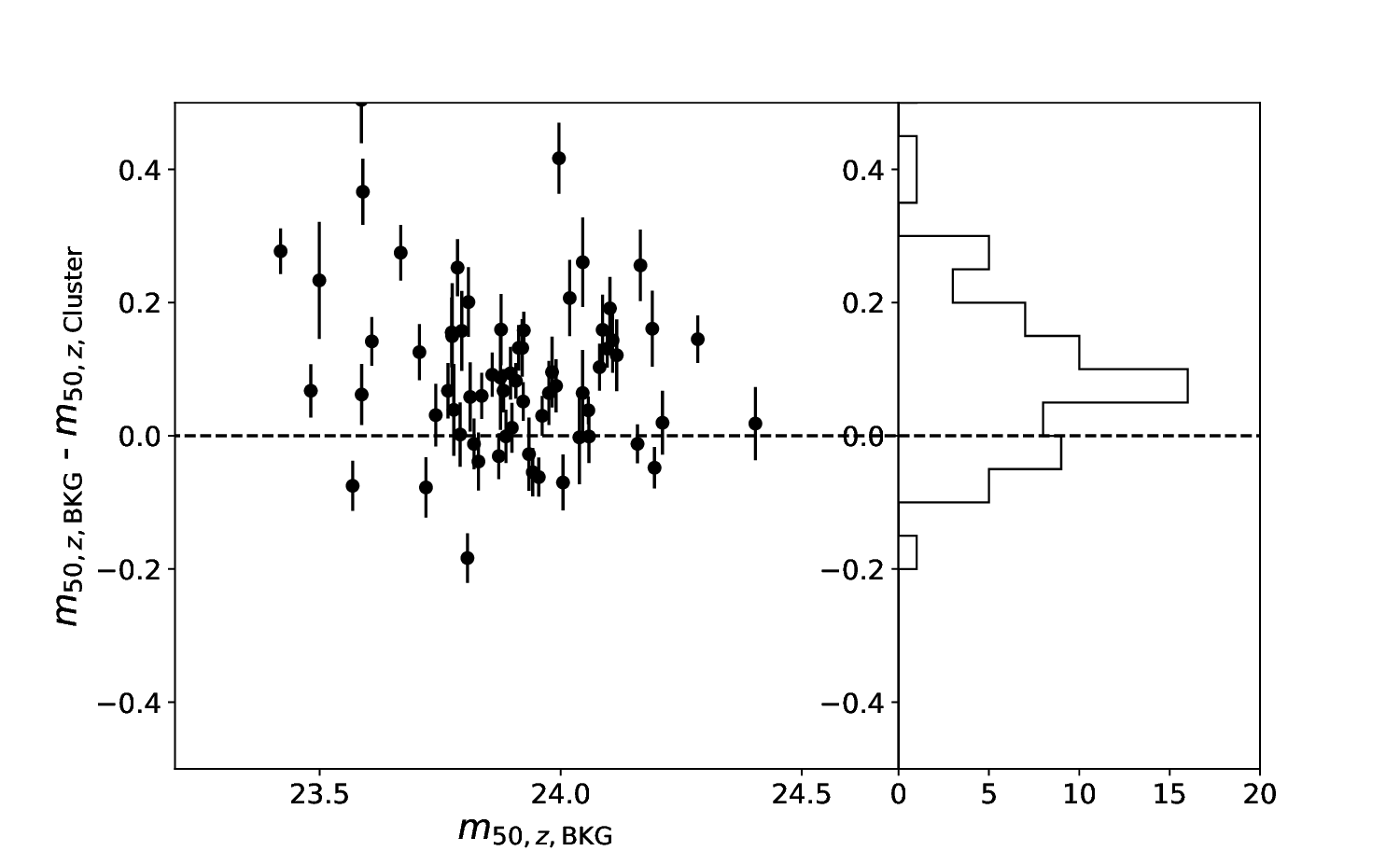}
\caption{We evaluate the $m_{50}$ parameters (50\% completeness magnitudes) for cluster and for field regions of the same depth with the \BALROG~simulation. The $m_{50}$ of a cluster region is potentially shallower by $\sim$ 0.1 mag compared to a same-depth field region potentially because of blending in the cluster region.}
\label{fig:rdmp_characterization}
\end{center}
\end{figure}

We notice hints that the completeness function in galaxy clusters are different from that of the fields, possibly because of blending and larger galaxy sizes. We test the effect with simulated objects \citep[\BALROG~simulation,][]{2015arXiv150708336S} inserted into RedMaPPer clusters \citep{2016ApJS..224....1R} selected in DES-SV data. 
We see evidence that the $m_{50}$ inside galaxy clusters shift by $\sim 0.1$ mag comparing to fields of equivalent depth (Figure~\ref{fig:rdmp_characterization}). As the sample of simulated galaxies is small, we are unable to characterize the distribution of the shifts and hence do not attempt to correct $m_{50}$ in this paper.

\subsection{Completeness Limits of the RSLF Analyses}

\begin{figure}
\begin{center}
\includegraphics[width=0.5\textwidth]{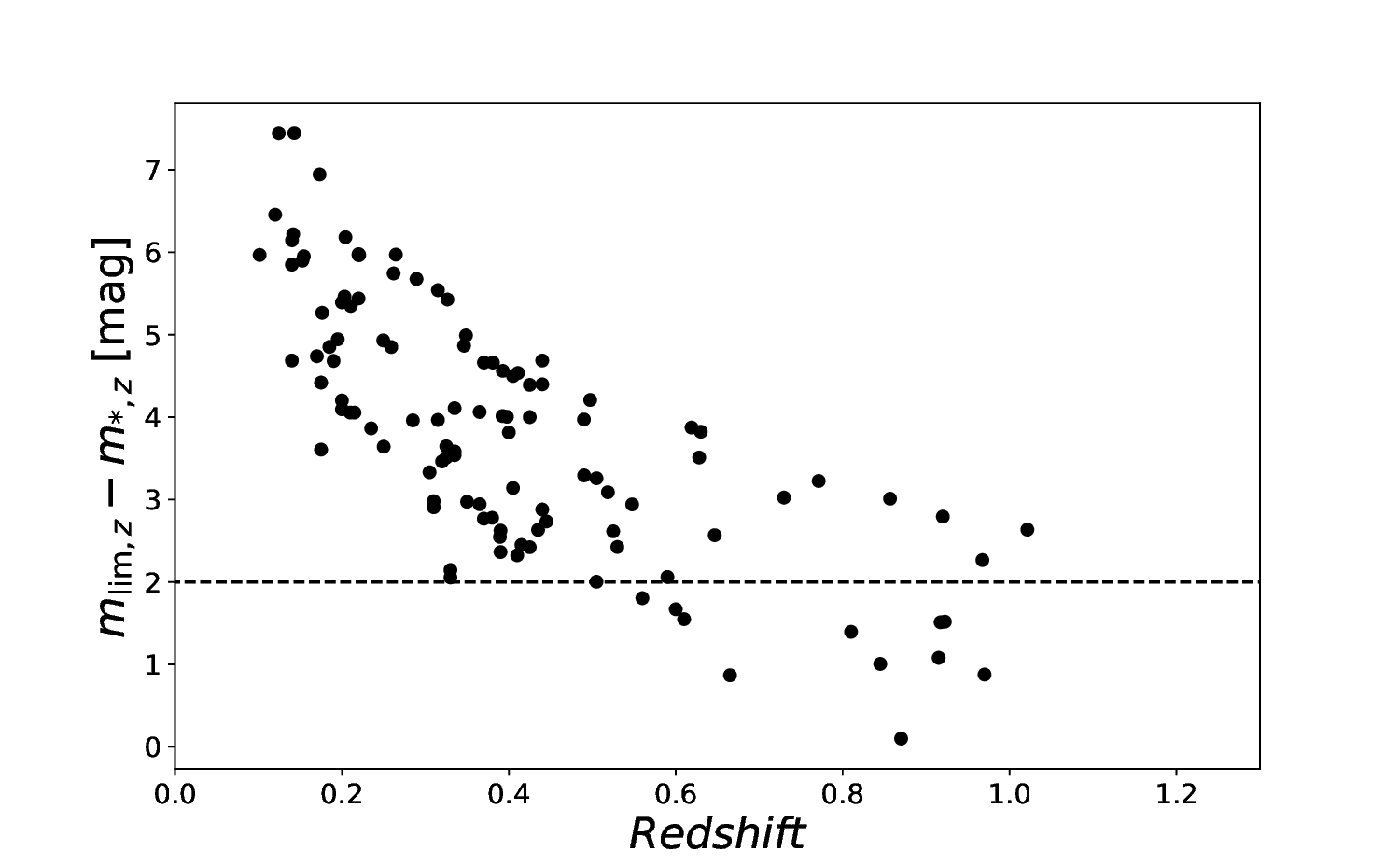}
\caption{For each cluster, we derive a completeness limit, $m_\mathrm{lim}$ from the completeness function. At $z<0.4$ , all of the DES XCS-SV clusters are complete to $m^*_z + 2~\mathrm{mag}$ and beyond. This is also true for more than $2/3$ of the clusters at $z>0.4$. Incomplete clusters of $m_\mathrm{lim}$ below $m^*_z + 2~\mathrm{mag}$ are not included in this paper's analyses. The scatters of $m_\mathrm{lim}$ are caused by DES depth variations in different parts of the sky.}
\label{fig:redshift_lim}
\end{center}
\end{figure}

We determine the magnitude limits of the RSLF analyses according to the completeness functions. We perform the analyses only with galaxies brighter than the following limit: $m_\mathrm{lim}=m_{50}-2\sqrt{2w}$. The cut ensures detection probability above 99.8\% $\times \lambda$ for the selected galaxies, according to our fitted completeness function model (Equation~\ref{eq:compl_func}). Note that the completeness limit is close to the 10-sigma total magnitude limit, which means galaxies above the completeness limit shall have total magnitude measured with significance level above or close to 10 $\sigma$, and hence above surface brightness detection limit set at the detection (1.5 sigma in SExtractor setup), and therefore any surface brightness selection effects should be negligible. 

If the cluster region completeness functions follow different relations as discussed above, the magnitude cut still ensures high detection probability (lower limit of 99\%$\times \lambda$ instead of 99.8\%$\times \lambda$). 

For all of the $z<0.4$ clusters, $m_\mathrm{lim}$ is more than 2 $\mathrm{mag}$ fainter than the characteristic magnitude measured in \cite{2009ApJ...699.1333H}. This is also true for more than $2/3$ of the clusters at $z>0.4$.  The cluster sample size drops steeply above redshift 0.7, and most of the complete clusters are located in the DES deep supernovae fields. As the galaxy samples are highly complete, we do not correct detection probability in this paper.

Because the $g$, $r$, $i$, $z$-band observations are performed independently, one may wonder if the image depth in the bluer bands is sufficient for  computing colors. For example, the $i$-band
 band observation of an object detected in $z$ may be too shallow that it does not have valid $i$-band photometry measurement. We confirm that after applying the z-band magnitude limit cut ($mag\_auto\_z < m_\mathrm{lim}$), 99.5\% and  99.6\% of the cluster region objects are detected in $r$ and $i$ respectively. 98.3\% or 99.2\% of the objects have good $r$ or $i$-band photometry measurement ($magerr\_auto$ above 3 $\sigma$, i.e., $magerr\_auto < 2.5/\mathrm{ln} 10/3$). We conclude that the DES multi-band data are sufficiently deep for red galaxy selection.

\section{Cluster Information}
\label{sec:sample_table}
%\Longtable
%\begin{deluxetable*}{cccccccc}
%\tablecaption{ \label{tbl:clusters}}
%\startdata
 \begin{table*}
 \label{tbl:clusters}
 \vspace{1em}
  \tiny
\begin{tabular}{lllll}
\hline
 Cluster Designation & R.A. & { \it Decl.} & $\mathrm{log} (M_{200}/M_\odot)$ & Cluster Redshift \\ 
 \hline
 XCSJ003248.5-431407.0 & 8.202084 & -43.235279 & 14.02 $\pm$ 0.16 & 0.3923$^a$ \\ 
  XCSJ003321.0-433737.1 & 8.337500 & -43.626972 & 14.08 $\pm$ 0.31 & 0.3809$^a$ \\ 
  XCSJ003346.3-431729.7 & 8.442917 & -43.291584 & 14.23 $\pm$ 0.12 & 0.2199$^a$ \\ 
  XCSJ003407.6-432236.2 & 8.531667 & -43.376720 & 13.89 $\pm$ 0.19 & 0.3928$^a$ \\ 
  XCSJ003428.0-431854.2 & 8.616667 & -43.315056 & 14.37 $\pm$ 0.10 & 0.3977$^a$ \\ 
  XCSJ003429.6-434715.7 & 8.623333 & -43.787693 & 13.47 $\pm$ 0.14 & 0.2042$^a$ \\ 
  XCSJ003518.1-433402.4 & 8.825417 & -43.567333 & 14.00 $\pm$ 0.12 & 0.4400$^a$ \\ 
  XCSJ003545.5-431756.0 & 8.939584 & -43.298889 & 13.48 $\pm$ 0.19 & 0.4109$^a$ \\ 
  XCSJ003548.1-432232.8 & 8.950417 & -43.375778 & 14.25 $\pm$ 0.16 & 0.6280$^a$ \\ 
  XCSJ003627.6-432830.3 & 9.115000 & -43.475082 & 13.94 $\pm$ 0.18 & 0.42 \\ 
  XCSJ004157.8-442026.5 & 10.490833 & -44.340694 & 13.94 $\pm$ 0.21 & 0.36 \\ 
  XCSJ021433.4-042909.9 & 33.639168 & -4.486084 & 14.77 $\pm$ 0.12 & 0.1401$^a$ \\ 
  XCSJ021441.2-043313.8 & 33.671665 & -4.553833 & 14.70 $\pm$ 0.11 & 0.1416$^a$ \\ 
  XCSJ021529.0-044052.8 & 33.870834 & -4.681334 & 14.31 $\pm$ 0.16 & 0.34 \\ 
  XCSJ021612.5-041426.2 & 34.052082 & -4.240611 & 14.32 $\pm$ 0.15 & 0.1543$^a$ \\ 
  XCSJ021653.2-041723.7 & 34.221668 & -4.289917 & 13.36 $\pm$ 0.17 & 0.1527$^a$ \\ 
  XCSJ021734.7-051327.6 & 34.394585 & -5.224333 & 14.08 $\pm$ 0.23 & 0.6467$^a$ \\ 
  XCSJ021741.6-045148.0 & 34.423332 & -4.863333 & 13.74 $\pm$ 0.26 & 0.5187$^a$ \\ 
  XCSJ021755.3-052708.0 & 34.480415 & -5.452222 & 13.74 $\pm$ 0.18 & 0.2495$^a$ \\ 
  XCSJ021803.4-055526.5 & 34.514168 & -5.924028 & 14.11 $\pm$ 0.26 & 0.3893$^a$ \\ 
  XCSJ021843.7-053257.6 & 34.682083 & -5.549333 & 13.73 $\pm$ 0.21 & 0.40 \\ 
  XCSJ021946.1-050748.2 & 34.942081 & -5.130055 & 14.44 $\pm$ 0.37 & 0.4902$^a$ \\ 
  XCSJ022024.7-050232.0 & 35.102917 & -5.042222 & 14.39 $\pm$ 0.16 & 0.12 \\ 
  XCSJ022034.4-054348.7 & 35.143333 & -5.730195 & 14.21 $\pm$ 0.26 & 0.20 \\ 
  XCSJ022042.7-052550.0 & 35.177917 & -5.430555 & 13.90 $\pm$ 0.21 & 0.5477$^a$ \\ 
  XCSJ022156.8-054521.9 & 35.486668 & -5.756083 & 14.08 $\pm$ 0.11 & 0.2591$^a$ \\ 
  XCSJ022204.5-043239.4 & 35.518749 & -4.544278 & 14.17 $\pm$ 0.29 & 0.3150$^a$ \\ 
  XCSJ022234.0-045759.8 & 35.641666 & -4.966611 & 13.77 $\pm$ 0.24 & 0.92 \\ 
  XCSJ022258.7-040637.9 & 35.744583 & -4.110528 & 14.01 $\pm$ 0.33 & 0.2893$^a$ \\ 
  XCSJ022307.9-041257.2 & 35.782917 & -4.215889 & 13.74 $\pm$ 0.31 & 0.6300$^a$ \\ 
  XCSJ022318.6-052708.2 & 35.827499 & -5.452278 & 13.61 $\pm$ 0.14 & 0.2106$^a$ \\ 
  XCSJ022342.3-050200.9 & 35.926250 & -5.033583 & 13.72 $\pm$ 0.18 & 0.8568$^a$ \\ 
  XCSJ022347.6-025127.1 & 35.948334 & -2.857528 & 14.10 $\pm$ 0.18 & 0.17 \\ 
  XCSJ022357.5-043520.7 & 35.989582 & -4.589083 & 14.16 $\pm$ 0.32 & 0.4974$^a$ \\ 
  XCSJ022401.9-050528.4 & 36.007915 & -5.091222 & 13.99 $\pm$ 0.15 & 0.3265$^a$ \\ 
  XCSJ022405.8-035505.5 & 36.024166 & -3.918195 & 14.32 $\pm$ 0.36 & 0.44 \\ 
  XCSJ022433.9-041432.7 & 36.141251 & -4.242417 & 13.91 $\pm$ 0.13 & 0.2619$^a$ \\ 
  XCSJ022457.9-034849.4 & 36.241249 & -3.813722 & 14.35 $\pm$ 0.15 & 0.6189$^a$ \\ 
  XCSJ022509.7-040137.9 & 36.290417 & -4.027194 & 13.89 $\pm$ 0.17 & 0.1732$^a$ \\ 
  XCSJ022512.2-062305.1 & 36.300835 & -6.384750 & 14.56 $\pm$ 0.15 & 0.2031$^a$ \\ 
  XCSJ022524.8-044043.4 & 36.353333 & -4.678722 & 14.26 $\pm$ 0.15 & 0.2647$^a$ \\ 
  XCSJ022530.8-041421.1 & 36.378334 & -4.239194 & 14.20 $\pm$ 0.12 & 0.1429$^a$ \\ 
  XCSJ022532.0-035509.5 & 36.383335 & -3.919306 & 14.30 $\pm$ 0.25 & 0.7712$^a$ \\ 
  XCSJ022808.6-053543.6 & 37.035831 & -5.595445 & 13.71 $\pm$ 0.13 & 0.21 \\ 
  XCSJ023037.2-045929.5 & 37.654999 & -4.991528 & 14.10 $\pm$ 0.21 & 0.31 \\ 
  XCSJ023052.4-045123.5 & 37.718334 & -4.856528 & 14.01 $\pm$ 0.15 & 0.31 \\ 
  XCSJ023142.2-045253.1 & 37.925835 & -4.881417 & 14.66 $\pm$ 0.11 & 0.20 \\ 
  XCSJ033150.1-273946.1 & 52.958752 & -27.662806 & 13.66 $\pm$ 0.18 & 1.0213$^a$ \\ 
  XCSJ034106.0-284132.2 & 55.275002 & -28.692278 & 14.60 $\pm$ 0.39 & 0.51 \\ 
  XCSJ041328.7-585844.3 & 63.369583 & -58.978973 & 13.64 $\pm$ 0.14 & 0.14 \\ 
  XCSJ041644.8-552506.6 & 64.186668 & -55.418499 & 14.24 $\pm$ 0.20 & 0.41 \\ 
  XCSJ042017.5-503153.9 & 65.072914 & -50.531639 & 14.17 $\pm$ 0.11 & 0.45 \\ 
  XCSJ043750.2-541940.8 & 69.459167 & -54.327999 & 13.83 $\pm$ 0.13 & 0.21 \\ 
  XCSJ043818.3-541916.5 & 69.576248 & -54.321251 & 14.94 $\pm$ 0.12 & 0.42 \\ 
  XCSJ065744.2-560817.0 & 104.434166 & -56.138054 & 14.14 $\pm$ 0.14 & 0.32 \\ 
  XCSJ065900.5-560927.5 & 104.752083 & -56.157639 & 14.08 $\pm$ 0.25 & 0.33 \\ 
  XCSJ095823.4+024850.9 & 149.597504 & 2.814139 & 14.56 $\pm$ 0.15 & 0.41 \\ 
  XCSJ095901.2+024740.4 & 149.755005 & 2.794556 & 13.90 $\pm$ 0.18 & 0.4900$^a$ \\ 
  XCSJ095902.7+025544.9 & 149.761246 & 2.929139 & 14.44 $\pm$ 0.13 & 0.3487$^a$ \\ 
  XCSJ095924.7+014614.1 & 149.852921 & 1.770583 & 13.96 $\pm$ 0.15 & 0.1243$^a$ \\ 
  XCSJ095932.1+022634.6 & 149.883743 & 2.442945 & 14.24 $\pm$ 0.25 & 0.42 \\ 
  XCSJ095940.7+023110.8 & 149.919586 & 2.519667 & 14.67 $\pm$ 0.15 & 0.7297$^a$ \\ 
  XCSJ095951.2+014045.8 & 149.963333 & 1.679389 & 14.11 $\pm$ 0.15 & 0.3702$^a$ \\ 
  XCSJ100023.1+022358.0 & 150.096252 & 2.399444 & 13.86 $\pm$ 0.13 & 0.22 \\ 
  XCSJ100027.1+022131.7 & 150.112915 & 2.358806 & 14.01 $\pm$ 0.16 & 0.2207$^a$ \\ 
  XCSJ100043.0+014559.2 & 150.179169 & 1.766444 & 14.30 $\pm$ 0.18 & 0.3464$^a$ \\ 
  XCSJ100047.3+013927.8 & 150.197083 & 1.657722 & 14.39 $\pm$ 0.11 & 0.2200$^a$ \\ 
  XCSJ224857.4-443013.6 & 342.239166 & -44.503777 & 15.08 $\pm$ 0.14 & 0.36 \\ 
  XCSJ232447.6-552443.3 & 351.198334 & -55.412029 & 13.91 $\pm$ 0.17 & 0.30 \\ 
  XCSJ232632.7-563054.5 & 351.636261 & -56.515141 & 13.73 $\pm$ 0.14 & 0.17 \\ 
  XCSJ232633.3-550116.3 & 351.638763 & -55.021194 & 14.41 $\pm$ 0.17 & 0.43 \\ 
  XCSJ232645.9-534839.3 & 351.691254 & -53.810917 & 13.55 $\pm$ 0.13 & 0.20 \\ 
  XCSJ232804.7-563004.5 & 352.019592 & -56.501251 & 14.15 $\pm$ 0.18 & 0.19 \\ 
  XCSJ232940.9-544715.3 & 352.420410 & -54.787582 & 13.71 $\pm$ 0.19 & 0.14 \\ 
  XCSJ232956.6-560808.0 & 352.485840 & -56.135555 & 14.32 $\pm$ 0.14 & 0.44 \\ 
  XCSJ233000.5-543706.3 & 352.502075 & -54.618416 & 14.34 $\pm$ 0.12 & 0.1763$^a$ \\ 
  XCSJ233037.2-554340.2 & 352.654999 & -55.727833 & 14.23 $\pm$ 0.28 & 0.33 \\ 
  XCSJ233132.2-531104.3 & 352.884155 & -53.184528 & 13.79 $\pm$ 0.17 & 0.41 \\ 
  XCSJ233133.8-562804.6 & 352.890839 & -56.467945 & 14.01 $\pm$ 0.29 & 0.18 \\ 
  XCSJ233204.9-551242.9 & 353.020416 & -55.211918 & 13.73 $\pm$ 0.16 & 0.34 \\ 
  XCSJ233216.0-544205.5 & 353.066681 & -54.701527 & 14.37 $\pm$ 0.19 & 0.32 \\ 
  XCSJ233225.7-560237.5 & 353.107086 & -56.043751 & 14.14 $\pm$ 0.18 & 0.28 \\ 
  XCSJ233403.8-554903.9 & 353.515839 & -55.817749 & 14.35 $\pm$ 0.35 & 0.34 \\ 
  XCSJ233706.9-541909.8 & 354.278748 & -54.319389 & 13.81 $\pm$ 0.31 & 0.53 \\ 
  XCSJ233835.2-543729.5 & 354.646667 & -54.624863 & 14.67 $\pm$ 0.28 & 0.38 \\ 
  XCSJ233955.1-561519.6 & 354.979584 & -56.255444 & 14.06 $\pm$ 0.39 & 0.37 \\ 
  XCSJ234054.4-554256.6 & 355.226654 & -55.715721 & 13.43 $\pm$ 0.15 & 0.17 \\ 
  XCSJ234142.9-555748.9 & 355.428741 & -55.963585 & 14.35 $\pm$ 0.16 & 0.20 \\ 
  XCSJ234231.5-562105.9 & 355.631256 & -56.351639 & 14.37 $\pm$ 0.14 & 0.35 \\ 
  XCSJ234311.1-555249.8 & 355.796265 & -55.880501 & 13.84 $\pm$ 0.20 & 0.23 \\ 
  XCSJ234600.9-561104.8 & 356.503754 & -56.184666 & 13.52 $\pm$ 0.14 & 0.1014$^a$ \\ 
  XCSJ234806.2-560121.1 & 357.025848 & -56.022530 & 14.79 $\pm$ 0.32 & 0.39 \\ 
  XCSJ235810.2-552550.1 & 359.542511 & -55.430584 & 14.57 $\pm$ 0.16 & 0.25 \\
  
 %\enddata 
$^a$ Archive spectroscopic redshift. 
%\end{deluxetable*}
\end{tabular}
 \end{table*}

\bibliographystyle{mnras}
\bibliography{references}

% Don't change these lines
\bsp	% typesetting comment
\label{lastpage}
\end{document}